\documentclass[usenatbib,referee]{mn2e}
\usepackage{epsf}
\usepackage{graphicx}

\def\eps@scaling{1.0}%
\newcommand\epsscale[1]{\gdef\eps@scaling{#1}}%
\newcommand\plotone[1]{%
 \centering
 \leavevmode
 \includegraphics[width={\eps@scaling\columnwidth}]{#1}%
}%

\def\mpc{\, h^{-1}{\rm {Mpc}}}

\def\msun{\, h^{-1}{\rm M_\odot}}
\def\mth{\, {\rm {M_{th}}}}

\def\gs{\mathrel{\raise1.16pt\hbox{$>$}\kern-7.0pt
\lower3.06pt\hbox{{$\scriptstyle \sim$}}}}
\def\ls{\mathrel{\raise1.16pt\hbox{$<$}\kern-7.0pt
\lower3.06pt\hbox{{$\scriptstyle \sim$}}}}
\def\gtsima{$\; \buildrel > \over \sim \;$}
\def\ltsima{$\; \buildrel < \over \sim \;$}
\def\prosima{$\; \buildrel \propto \over \sim \;$}
\def\gsim{\lower.5ex\hbox{\gtsima}}
\def\lsim{\lower.5ex\hbox{\ltsima}}
\def\simgt{\lower.5ex\hbox{\gtsima}}
\def\simlt{\lower.5ex\hbox{\ltsima}}
\def\simpr{\lower.5ex\hbox{\prosima}}
\def\la{\lsim}
\def\ga{\gsim}

\title[Halos and the environment] {Internal properties and environments of dark matter
halos}

\author[Huiyuan Wang et al.]
   {\parbox[t]{\textwidth}{
       Huiyuan Wang$^{1,2}$\thanks{E-mail: whywang@mail.ustc.edu.cn},
       H. J. Mo$^{3}$,
       Y.P. Jing$^{4}$,
       Xiaohu Yang$^{4}$
       Yu Wang$^{1,2}$
}\\
            $^1$Key Laboratory for Research in Galaxies and
            Cosmology, University of Science and Technology of China, Hefei, Anhui 230026, China\\
            $^2$Department of Astronomy, University of Science and Technology of China,
            Hefei, Anhui 230026, China\\
            $^3$Department of Astronomy, University of Massachusetts,
            Amherst MA 01003-9305, USA\\
            $^4$Key Laboratory for Research in Galaxies and Cosmology, Shanghai Astronomical Observatory, Shanghai 200030,
China }

\date{
Accepted ........ Received .......; in original form ......}

\pubyear{2010}

\begin{document}

\maketitle \label{firstpage}

\begin{abstract}
We use seven high-resolution $N$-body simulations to study the
correlations among different halo properties (assembly time, spin,
shape and substructure), and how these halo properties are
correlated with the large-scale environment in which halos reside.
The large-scale tidal field estimated from halos above a mass
threshold is used as our primary quantity to characterize the
large-scale environment, while other parameters, such as the local
overdensity and the morphology of large-scale structure, are used
for comparison. For halos at a fixed mass, all the halo properties
depend significantly on environment, particularly the tidal field.
The environmental dependence of halo assembly time is primarily
driven by local tidal field. The mass of the unbound fraction in
substructure is boosted in strong tidal force region, while the
bound fraction is suppressed. Halos have a tendency to spin faster
in stronger tidal field and the trend is stronger for more massive
halos. The spin vectors show significant alignment with the
intermediate axis of the tidal field, as expected from the tidal
torque theory. Both the major and minor axes of halos are strongly
aligned with the corresponding principal axes of the tidal field.
In general, a halo that can accrete more material after the
formation of its main halo on average is younger, is more
elongated, spins faster, and contains a larger amount of
substructure. Higher density environments not only provide more
material for halo to accrete, but also are places of stronger
tidal field that tends to suppress halo accretion. The
environmental dependencies are the results of these two competing
effects. The tidal field based on halos can be estimated from
observation, and we discuss the implications of our results for
the environmental dependence of galaxy properties.
\end{abstract}

\begin{keywords}
dark matter - large-scale structure of Universe - galaxies: halos
- methods: statistical
\end{keywords}

\section{Introduction}
In the cold dark matter cosmogony, a  key  concept in the build-up
of cosmic structure is the formation of dark matter halos. These
halos are not only the building blocks of the large-scale
structure of the Universe, but also the hosts within which
galaxies are supposed to form. During the last decade, the
properties of the dark halo population, such as their internal
structures, kinematic properties, assembly histories and
clustering properties, have been studied in great detail using
both numerical and analytical methods. The results obtained have
provided important clues about the formation and evolution of
galaxies in the cosmic density field. For example, the dependence
of halo clustering on mass (e.g., Mo \& White 1996; Jing 1998;
Sheth \& Tormen 1999; Sheth, Mo \& Tormen 2001; Seljak \& Warren
2004), referred to as the halo bias, has widely been used to
interpret the clustering properties of galaxies via the halo
occupation model (e.g., Jing, Mo \& B\"{o}rner 1998; Peacock \&
Smith 2000) and the conditional luminosity function model (e.g.,
Yang, Mo \& van den Bosch 2003); the halo shape and orientation
have offered useful constraints on the formation and evolution of
galaxy clusters (e.g. Jing \& Suto 2002; Hopkins et al. 2005); the
angular momentum properties of dark matter halos have played a
crucial role in the modelling of the formation of disk galaxies
(e.g. Fall \& Efstathiou 1980; Mo, Mao \& White 1998); and the
assembly histories of halos have played an important role in the
understanding of star formation and morphology of galaxies (e.g.
Mo \& Mao 2004; Dutton et al. 2007; van den Bosch 2002).

The acquisition of angular momentum of dark matter halos has been
studied for a long time. Halo spin is thought to be generated by
the tidal torques exerted by large scale structure (e.g. Peebles
1969; White 1984). The tidal torque theory successfully reproduces
the characteristic spin distribution of halos, although the
theoretical prediction (based on quasi-linear theory) of the
alignment between the spin axis and the tidal field is not
detected in N-body simulations (Porciani, Dekel, \& Hoffman 2002).
Alternatively, a number of studies have considered the possibility
of generating halo angular momentum through mergers (Gardner 2001;
Vitvitska et al. 2002; Maller, Dekel \& Somerville 2002;
Hetznecker \& Burkert 2006). In particular, Maller et al. (2002)
found that the spin distribution seen in cosmological N-body
simulations can be reproduced by the merger scenario. Clearly, the
origin of halo angular momentum remains an unresolved problem, and
detailed analysis of the correlation between halo spins and other
halo and environmental properties is required to shed light on it.

The existence of substructure (subhalos) within dark matter halos
is a natural consequence of hierarchical structure formation, and
the properties of the subhalo population have been studied
extensively using N-body simulations (e.g. Moore et al. 1999;
Klypin et al. 1999; Gao et al. 2004; Diemand, Kuhlen \& Madau
2007). Recently, Ludlow et al. (2009) extended the analysis by
examining all halos physically associated with host halos. Some of
the haloes were found to be once inside host haloes and
subsequently ejected (see also Wang et al. 2009b). This suggests
that some fraction of the subhalos currently residing in their
hosts may not be bound to the hosts and will eventually escape.
The presence of this unbound subhalo population may affect the
properties derived for the host halos. Indeed, D'Onghia \& Navarro
(2007) found that the ejection of high-angular momentum material
can reduce the spin of a halo that has ceased growing.
Understanding the correlations between these unbound substructures
with other halo properties can thus provide insight into the
formation of dark matter halos.

More recently, a number of independent investigations based on
high-resolution N-body simulations have found that the clustering
strength of halos of fixed mass depends significantly on other
halo properties, such as assembly time, substructure, spin, shape
and concentration (e.g. Sheth \& Tormen 2004; Gao et al. 2005;
Harker et al. 2006; Wechsler et al. 2006; Jing, Suto \& Mo 2007;
Hahn et al. 2007a; Wetzel et al. 2007; Bett et al. 2007; Gao \&
White 2007; Li et al. 2008). Such dependencies, sometimes referred
together as the assembly bias, indicate that the formation of
halos may be affected by large-scale environmental effects other
than what produces the halo bias (the correlation between halo
mass and large-scale environment). Since these properties of halos
may be related to the properties of the galaxies they host, an
understanding of these effects can help us to understand how
galaxies of different properties form and reside in the cosmic
density field. (Yang, Mo \& van den Bosch 2006a; Zhu et al. 2006;
Wang et al. 2007b, 2008a; Croton, Gao, \& White 2007; Tinker et
al. 2008; Bamford et al. 2009; Skibba \& Sheth 2009; Weinmann et
al. 2009). In addition, observations also show that the spin
axes/orientations of galaxies tend to align with galaxy
distribution in the neighborhood (e.g. Holmberg 1969; Binggeli
1982; Yang et al. 2006b; Lee \& Erdogdu 2007; Faltenbacher et al.
2009), suggesting again that the spins and orientations of dark
matter halos are correlated with large-scale structure.

There is a large body of theoretical investigations about the
origins of the correlation between halo properties and large-scale
environments, especially the origin of the halo assembly bias
(Wang, Mo \& Jing 2007a; Zentner 2007; Sandvik et al. 2007;
Keselman \& Nusser 2007; Desjacques 2008; Dalal et al. 2008; Hahn
et al. 2009; Wang, Mo \& Jing 2009b; Fakhouri \& Ma 2009). Most of
these studies focussed on the environmental dependence of halo
\emph{assembly time}. The results suggested that the growth of old
small halos can be suppressed by the tidal field induced by nearby
massive structures, and an assembly bias can be produced through
the tidal truncation of the growth of small halos in high-density
regions. However, as pointed out by Gao \& White (2007), the
environmental dependencies differ qualitatively for different halo
properties. In particular, the environmental dependencies of the
assembly time and substructure fraction appear to be inconsistent
with the correlation between the assembly time and substructure
fraction. Furthermore, the environmental dependencies of halo
shape and spin also seem to be in conflict with the correlation
between them (see Bett et al. 2007). Thus, the assembly-time
effect alone may not be able to explain all the correlations seen
in simulations.

In order to improve our understanding about how environmental
effects affect the formation and structure of dark matter halos,
it is important to gather more information about the halo -
environment connection from large N-body simulations. In this
paper, we carry out a systematic analysis of the correlation
between various halo properties and environments. Our analysis
differs from earlier investigations in the following two aspects.
First, we introduce a new method to quantify the large-scale
environments of halos. We use the large scale tidal field
\emph{estimated from a population of halos above a certain mass
threshold} as our primary environment indicator. As we will show
below, this quantity is more strongly correlated with halo
properties than other environment indicators, such as the local
density and morphology of large-scale structure. Second, we divide
the substructures into two components according to whether or not
the mass is bound and unbound to the host halo. As we will see
later, this allows us to separate different environmental effects
and understand why some environmental dependencies of halo
properties presented in the literature appear to be inconsistent
with each other.

The paper is organized as follows. In Section \ref{sec_sim}, we
describe the simulations to be used, how dark halos are
identified, and the methods to compute various halo properties. We
then show the correlations among different halo properties in
Section \ref{sec_chp}. In Section \ref{sec_env}, we describe our
method to quantify the environments of dark matter halos. We
analyze the correlations between halo properties and environments
in Section \ref{sec_che} to find out the processes that affect the
properties of halos. Finally, in Section \ref{sec_con}, we
summarize our results and discuss their implication.

\section{Simulations and Dark Matter halos}\label{sec_sim}

\subsection{Simulations}

In this paper, we use seven $N$-body simulations and dark matter
halos  selected  from them to study the correlations of halo
properties with the large scale environment. These simulations are
obtained using the ${\rm P^3M}$ code described in Jing et al.
(2007). Three of them, which will be referred to as L300, assume a
spatially-flat $\Lambda$CDM model, with the density parameter
$\Omega_{\rm m}=0.268$, the cosmological constant
$\Omega_\Lambda=0.732$ and the baryon density parameter
$\Omega_b=0.045$, and with the $\Lambda$CDM power spectrum
obtained from CMBfast (Seljak \& Zaldarriaga 1996) with an
amplitude specified by $\sigma_8=0.85$. The CDM density field of
each simulation was traced by $1024^3$ particles, each having a
mass of $M_p\sim1.87 \times 10^{9}\msun$, in a cubic box of 300
$\mpc$. The other four simulations, referred to as L150 in the
following, assume the same cosmological model as L300, and use the
same number of particles, but the simulation box is smaller,
150$\mpc$, and the mass resolution is higher,
$M_p\sim2.34\times10^{8}\msun$.

Dark matter halos were identified using the standard
friends-of-friends algorithm (e.g. Davis et al. 1985) with a link
length that is 0.2 times the mean inter-particle separation. The
mass of a halo, $M_h$, is the sum of the masses of all the
particles in the halo. The virial radius $R_h$ of a halo is
defined as:
\begin{equation}
R_h=\left({3 M_h\over 4\pi \Delta_h{\rho_{\rm m}}}\right)^{1/3}\,,
\label{eq_rvir}
\end{equation}
where  $\rho_{\rm m}$  is  the  mean mass   density  of  the
universe, and $\Delta_h$ is the  mean  density contrast  of  a
virialized  halo chosen to be $\Delta_h = 200$ (e.g. Porciani,
Dekel \& Hoffman 2002).

\subsection{Halo assembly times}

Halos at $z = 0$ are linked to their progenitors at higher $z$
through halo merger trees (e.g. Lacey \& Cole 1993). A halo in an
earlier output is considered to be a progenitor of the present
halo if more than half of its particles are found in the present
halo. The assembly time of the halo, $z_{\rm f}$, is defined as
the redshift at which the most massive progenitor first reaches
half of the final mass of the halo. Interpolations between
adjacent outputs were adopted when estimating $z_{\rm f}$. In
order to get a reliable estimate of the assembly time of a halo
from the simulation, one needs to follow the growth of the main
progenitor accurately. A halo can be identified reliably if it
contains more than 50 particles (e.g. Gao et al. 2005). We thus
only use halos that contain more than 100 particles at $z = 0$.

\subsection{Mass fraction in halo substructure}

We use SUBFIND developed by Springel et al. (2001) to identify
substructures within an FOF halo. Each halo is decomposed into a
set of self-bound subhalos, down to 10 particles, plus a `fuzz' of
unbound particles which contains unresolved subhalos. The most
massive subhalo is referred to as the main halo of the FOF halo,
and the rest subhalos and fuzz are referred to as substructure.
Following Gao \& White (2007), we use the parameter $f_{\rm
s}=1-M_{\rm main}/M_h$ to describe the amount of substructure,
where $M_{\rm main}$ and $M_h$ are the masses of the main halo and
the FOF halo, respectively. SUBFIND identifies the main halo after
the removal of all particles in the substructure, regardless
whether or not the particles are bound to the main halo. The
substructure fraction, $f_{\rm s}$, therefore contains two
components: bound and unbound. We calculate the energy (the sum of
kinetic energy and potential energy) of subhalos relative to the
main halo. If the energy is negative, the subhalo is said to be
bound to the main halo, else it is not. We thus can define another
two parameters, $f_{\rm us}$ and $f_{\rm bs}$, with $f_{\rm bs}$
the mass ratio of bound subhalos to the FOF halo, and $f_{\rm us}$
that of unbound subhalos and fuzz. Clearly, $f_{\rm bs}+f_{\rm
us}= f_{\rm s}$. The substructure fraction $f_{\rm s}$ so defined
is a measure of the mass a halo assembles after the formation of
its main halo, while $f_{\rm us}$ versus $f_{\rm bs}$ indicates
what kind of mass is assembled.

The measurement of $f_{\rm bs}$ from simulation can be affected by
mass resolution, because some small subhalos are unresolved. Here
we give a rough estimation of this effect. It is well known that
the cumulative subhalo mass function is well-fit by a power-law
$N(>M_s)=A (M_s/M_h)^{-1}$ down to $M_s\simeq 10^{-5}M_h$(e.g.
Diemand et al. 2007), where $M_s$ is the subhalo mass and the
amplitude $A$ depends on halo properties (e.g. Moore et al. 1999;
Gao et al. 2004). Therefore $f_{\rm bs}(M_{sc}/M_h)\simeq
A\ln(M_h/M_{sc})=A\ln(N_h/10)$, where $N_h$ is the number of dark
matter particles within the halo, and $M_{sc}= 10 M_p$ is the mass
of the smallest subhalos identified. Clearly, the substructure
mass fraction to be measured is sensitive to the particle number
contained in the host. To reduce this effect, we divide halos into
several narrow mass bins with size of 0.5 dex. In each mass bin,
the variation of $f_{\rm bs}$ due to mass resolution is $\log
(0.5\ln 10) \simeq 0.06$ dex, much less than the intrinsic scatter
(more than 1 dex, see figure 8 of Gao et al. 2004). On the other
hand, in order to make sure that the measured $f_{\rm bs}$ is not
dominated by statistical variation, $M_{sc}/M_h$ has to be
sufficiently small. As adopting $M_{sc}/M_h \leq 0.3\%$, more than
half of mass in subhalos of $M_s\geq10^{-5}M_h$ can be identified
and the statistical variation is thus small. This suggests that
$f_{\rm bs}$ can be reliably determined for halos containing more
than 3300 particles. The mass resolution does not have a
significant effect on the parameter $f_{\rm us}$. If the
unresolved subhalos are not bound to the main halo, they must be
contained in the fuzz. Therefore the unresolved subhalos still
contribute to $f_{\rm us}$ without being missed. The uncertainty
in $f_{\rm us}$ is expected to be less than $1/N_h$. For $N_h >
3300$, it is much smaller than the typical value of $f_{\rm us}$
(see the following).

\subsection{Halo spin parameters}

 The importance of the rotational motion relative
to the internal random motion within a halo is usually described
by a dimensionless spin parameter (Peebles 1969), defined as,
\begin{equation}
\lambda=\frac{J|E|^{1/2}}{GM_h^{5/2}}
\end{equation}
where $G$ is Newton's gravitational constant, $E$ is the total
energy and $J$ the magnitude of angular momentum of the halo. We
calculate the total energy using the method described in Bett et
al. (2007). If a FOF halo contains more than 4000 particles, the
total energy is computed using a random set of 4000 particles;
otherwise all particles are used. We have measured the angular
momentum, $J$, relative to the mass center whose velocity is
defined by an average over all particles contained in the halo. We
use $\hat {j}$, the direction of halo angular momentum, to denote
the rotational axis of the halo.

\subsection{Halo shapes and orientations}

We use the inertia momentum tensor, ${\cal I}$, of a halo to
characterize its shape and orientation (Jing et al. 1995). The
components of ${\cal I}$ are estimated using
\begin{equation}
I_{jk}=M_p\sum^N_{n=1}x_{n,j}x_{n,k},
\end{equation}
where $x_{n,j}$ ($j=1, 2$ or 3) are the components of the position
vector of the $n$th particle relative to the center of mass of the
halo. The square root of the eigenvalues of this inertia momentum
tensor are often used to represent the principal axes, $I_1$,
$I_2$ and $I_3$ ($I_1\geq I_2\geq I_3$). In this paper, we use the
axis ratios, e.g. $I_3/I_1$, to characterize the shape of a halo,
and the corresponding unit vectors, $\hat{I}_1$, $\hat{I}_2$ and
$\hat{I}_3$ to denote the directions of the major, intermediate
and minor axes, respectively.

 The estimates of both the spin parameter and the axial ratio
can be affected significantly by numerical resolutions. As shown
in Bett et al. (2007), in order to estimate these two parameters
reliably, one needs more than 300 particles to sample a halo. To
achieve such mass resolution and at the same time to obtain
sufficient number of halos to perform the statistical analysis, we
use the simulation sets L150 to study halos with masses $12<\log
(M_h/\msun)<13$ (so that each halo contains more than 4270
particles) and use L300 to study halos with $\log (M_h/\msun)>13$
(more than 5300 particles per halo). Since the environmental
effects on halo assembly time is strong only for low-mass halos
(see below), we will include halos with $10.7<\log (M_h/\msun)<11$
(more than 210 particles per halo) in the L150 simulations when
assembly time is considered. In most case, we do not consider
halos of with $\log (M_h/\msun)>14.5$, because the total number of
halos in this mass range is too small to give statistically
reliable results.

\section{Correlations among halo properties}\label{sec_chp}

\begin{figure}
\epsscale{1}\plotone{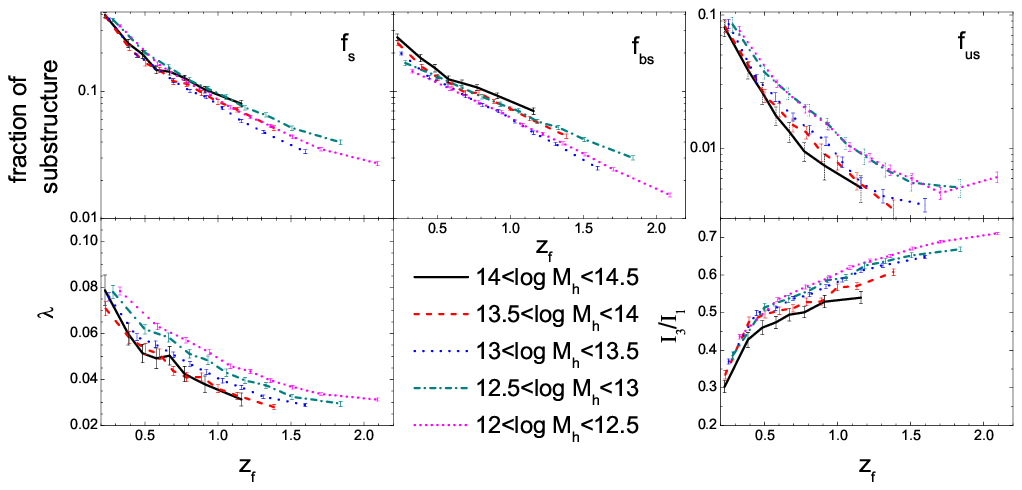}
 \caption{Median values of the
fraction of total substructure, $f_{\rm s}$, fraction of bound
substructure, $f_{\rm bs}$, fraction of unbound substructure,
$f_{\rm us}$, spin parameter, $\lambda$, and short-to-long axis
ratio, $I_3/I_1$, as functions of halo assembly time, $z_{\rm f}$,
for halos of various masses. The halo mass range for each case is
shown in the lower-middle panel.} \label{fig_zfhp}
\end{figure}

We first examine how halo properties, such as $f_{\rm s}$, $f_{\rm
us}$, $f_{\rm bs}$, $\lambda$, and $I_3/I_1$, are correlated with
halo assembly time (specified by $z_{\rm f}$). The median values
of these parameters as functions of $z_{\rm f}$ for various narrow
mass bins are plotted in Fig. \ref{fig_zfhp}. The errors, $e_y$,
on the median of a parameter, $y$, shown in the figures are
computed using
\begin{equation}
e_y=\frac{y_{84}-y_{16}}{\sqrt{N_{halo}}}\,,
\end{equation}
where $N_{halo}$ is the number of halos in each $z_{\rm f}$ bin
(note that bin sizes are chosen so that each bin contains an equal
number of halos), $y_{84}$ and $y_{16}$ denote the 84th and 16th
percentiles of the distribution of $y$, corresponding to a
$1\sigma$ spread if the underlying distribution were Gaussian.

All halo properties considered here show significant correlation
with the assembly time. On average, young halos (those with lower
$z_{\rm f}$) contain more substructures, spin more rapidly and are
less spherical, than old halos of the same mass. These results are
consistent with those found before (e.g. Jing \& Suto 2002; Gao et
al. 2004; Allgood et al. 2006; Hahn et al. 2007a). Many authors
have interpreted these correlations as due to the fact that newly
accreted halos may survive in their host halos (Gao et a. 2004),
so as to significantly enhance the spin of the hosts (e.g.
Vitvitska et al. 2002; Hetznecker \& Burkert 2006), and to make
the hosts more elongated (e.g. Hopkins et al. 2005). Such
interpretation is supported by the fact that halo spin and shape
have stronger correlation with the substructure fraction than with
the assembly time (see Fig. \ref{fig_fslh} below). Furthermore, we
find that the $z_f$-dependence of the short-to-intermediate axial
ratio, $I_3/I_2$, is much weaker than the other two ratios,
indicating that new material tends to be accreted along the major
axes of halos (Wang et al. 2005).

However, other processes may also be important, at least for some
of these correlations. For instance, D'Onghia \& Navarro (2007)
found that the spins of halos that have ceased growing can still
drop gradually, presumably due to mass redistribution such as the
ejection of high-angular momentum material from the halo during
the subsequent virialization process. Indeed, as shown in Wang et
al. (2009b), there is clear evidence that old halos tend to eject
more subhalos than young ones of the same mass. A correlation
between halo spin and assembly time can thus be produced via this
process. Clear, more detailed analysis are needed to quantify the
role of such mechanism.

Fig. \ref{fig_fslh} shows how halo spin and axis ratio are
correlated with the substructure fractions. Both $\lambda$ and
$I_3/I_1$ depend strongly on unbound and bound fractions. This is
consistent with Maccio' et al. (2007; see also Shaw et al. 2006)
who found that unrelaxed halos tend to spin more rapidly and to be
more prolate. As one can see from the left panel of
Fig.\ref{fig_js}, which shows the correlation between halo spin
and halo axis ratio, less spherical halos, especially the ones
with low masses, tend to have higher $\lambda$. In the right panel
of the figure, we show the probability distribution function of
the cosine of the angle between the spin vector, $\hat{j}$, and
the three principle axes of the halo, $\hat{I}_1$, $\hat{I}_2$ and
$\hat{I}_3$. Since we do not find any evidence for such alignment
to depend on halo mass, results are shown only for two mass bins.
As one can see, the spin axis has the tendency to be parallel to
the minor axis and perpendicular to the major axes (see also
Warren et al. 1992; Shaw et al. 2006; Bett et al. 2007; Zhang et
al. 2009).

 Inspecting these correlations in detail, we can find
several interesting trends. First, the dependence of halo spin on
$f_{\rm us}$ is stronger than on any other halo properties, such
as $z_{\rm f}$ and $f_{\rm bs}$. In contrast, the halo axial ratio
shows stronger correlation with $z_{\rm f}$ and $f_{\rm bs}$ than
with $f_{\rm us}$. Second, the relationships among $f_{\rm bs}$,
$z_{\rm f}$ and $I_3/I_1$ (see the upper-middle and lower-right
panels of Fig. \ref{fig_zfhp}, and the lower-middle panel of Fig.
\ref{fig_fslh}) and between $f_{\rm us}$ and $\lambda$ (see the
upper-right panel of Fig.\,\ref{fig_fslh}) have only weak
dependencies on halo mass, while the relationships between these
two sets of halo properties depend significantly on halo mass.
These tendencies suggest that $f_{\rm bs}$, $z_{\rm f}$ and
$I_3/I_1$ may have similar origin, and so do $f_{\rm us}$ and
$\lambda$. In Section \ref{sec_che} we will investigate how these
halo properties depend on halo environment, and we will see that
$f_{\rm bs}$, $z_{\rm f}$ and $I_3/I_1$ exhibit similar
environmental dependence, while $f_{\rm us}$ and $\lambda$ exhibit
common environmental dependence different from that of the other
three halo properties.

\section{Large-scale tidal field traced by halos}\label{sec_env}

 In the literature a number of parameters have been used
to quantify the large-scale environments. These include the local
mass overdensity around halos, the halo bias parameter, the
morphology of large scale structure, and the tidal field produced
by large-scale mass distribution. In this paper we use the `halo
tidal field', obtained from the distribution of dark matter halos
above a certain mass threshold $\mth$, as our primary
environmental parameters. As shown in Yang et al. (2005; 2007),
galaxy groups/clusters properly selected from large redshift
surveys of galaxies can be used to represent the dark halo
population, especially massive ones with masses $\ga
10^{12}\msun$. Thus, the `halo tidal field', can in principle be
estimated from observation. In what follows we describe how the
halo tidal field is defined and estimated. In Appendix B we
examine how halo tidal field is correlated with the other
environmental indicators mentioned above.

The normalized halo tidal force \emph{on the surface} of a given
halo, `h', in a direction $\vec{t}$ is defined as
\begin{equation}
F_t(\vec{t})= \frac{\sum_{i=1}^N\frac{GM_{i}}{r_i^3}
R_h(1/2+3/2\cos(2\theta_i))}{GM_h/R_h^2}
=\sum_{i=1}^{N}\frac{R_{i}^3}{2r_i^3}(1+3\cos{2\theta_i})\,,
\label{eq_ft}
\end{equation}
where $M_h$ and $R_h$ are the mass and radius of the halo in
question, $M_{i}$ and $R_{i}$ are the masses and radii of other
halos producing the tidal force,  $r_i$ is the distance from halo
`$h$' to halo `$i$', and $\theta_i$ is the angle between $\vec{t}$
and $\vec{r}_i$. The second equation follows from the fact that
the mean density within the virial radius at a given redshift is
the same for all halos, so that $M_h\propto R_h^3$ and $M_i\propto
R_i^3$. Thus, the tidal force on a halo is calculated by summing
up the tidal forces of all other halos of mass above $\mth$, and
is normalized by the self-gravity of the halo in question, so that
one can compare the environmental effects for halos of different
masses. We define the halo tidal force on the halo surface so that
we can easily quantify/distinguish between the self-gravity or
tidal force dominated impact on the particles that are to be
accreted to or to be ejected from the halos. In the following, we
adopt a threshold mass $\mth=10^{12}\msun$, which is the low mass
limit of groups selected from current galaxy redshift surveys
(e.g. Yang et al. 2007). And as we have tested, using somewhat
larger or smaller $\mth$ does not change any of our results
significantly.

We first define two tidal directions, $\vec{t}_1$ and $\vec{t}_3$,
so that the tidal force has the largest value along $\vec{t}_1$
and the lowest value along $\vec{t}_3$. According to the analysis
presented in the Appendix, vectors $\vec{t}_1$ and $\vec{t}_3$ are
the eigenvectors of the halo tidal tensor, are perpendicular to
each other, and represent the major and minor axes of the halo
tidal field. The third tidal direction, $\vec{t}_2$, is defined as
a vector perpendicular to both $\vec{t}_1$ and $\vec{t}_3$. We use
$t_1$, $t_2$ and $t_3$ to denote the tidal forces along
$\vec{t}_1$, $\vec{t}_2$ and $\vec{t}_3$, respectively. Different
from the tidal field produced by large-scale mass distribution,
halo tidal field satisfies $t_1+t_2+t_3=0$ (see Appendix A); thus
only two parameters are needed to characterize the halo tidal
field. We adopt $t_1$ to represent the magnitude and a parameter,
\begin{equation}
t_s\equiv \frac{t_1-t_2}{t_1+t_2}\,,
\end{equation}
to characterize the `shape' of the tidal field. Clearly, $t_s$
describes the anisotropy in the distribution of neighboring halos.
If $t_s>1$, then both $t_2$ and $t_3$ must be negative while
$t_1>0$. Thus the tidal field stretches the material along
$\vec{t}_1$, but compresses it in the other two directions. This
also means that halos dominating the tidal field must be
distributed preferentially along $\vec{t}_1$ in a filamentary
structure. In the extreme case of $t_s=3$, the tidal field is
dominated by just one halo. If $t_s<1$, then $t_2>0$ so that the
tidal field compresses halo material only along $\vec{t}_3$, while
stretches it along both $\vec{t}_1$ and $\vec{t}_2$. Such a tidal
field can be produced by more than one halos distributed
preferentially in the $\vec{t}_1$-$\vec{t}_2$ plane.

\section{Correlations between halo properties and environment}\label{sec_che}

In this section, we explore the correlations between halo
properties and their environments. The goal is to find out which
environmental effects have the strongest impact on halo
properties, and whether there is any connection between different
environmental effects. Since halo mass may be correlated with both
environment and other halo properties, it is sometimes necessary
to divide halos into narrow mass bins in order to identify
possible causal connections between halo properties and
environment. In such cases we divide halos into mass bins with a
size of 0.5 dex or smaller.

\subsection{Assembly time}\label{sec_at}

 It has been found that the clustering of low-mass halos
depends on their assembly times (e.g. Gao et al. 2005). Wang et
al. (2007a) found that old, low-mass halos have a tendency to
reside in the vicinity of massive systems, and suggested that the
tidal truncation of accretion may be responsible for the assembly
bias (see also Keselman \& Nusser 2007; Desjacques 2008). In Fig.
\ref{fig_zft}, we show the relation between halo assembly time,
$z_{\rm f}$, and the halo tidal field represented by $t_1$. As one
can see, the median assembly time increases with $t_1$ for
low-mass halos, and the dependence becomes weaker with the
increase of halo mass.

Since $t_1$ is correlated with the large-scale density field and
the morphology of large scale structure (see Appendix B), the
dependence of assembly time on tidal field might originate from
the correlation between assembly time with these other environment
quantities. Clearly, further analysis is required in order to
examine which quantity plays the more primary role in affecting
halo assembly. Since strong environmental dependence of $z_{\rm
f}$ exists only for low-mass halos, we concentrate on halos with
$10.7<\log (M_h/\msun)<11$. The upper-left panel of Fig.
\ref{fig_zfe} shows the median assembly time versus $1+\delta
(6\mpc)$, where $\delta (6\mpc)$ is the overdensity of dark matter
within a sphere of radius 6 $\mpc$ around each dark matter halo
(defined in Appendix B). It is evident that halos in higher
density regions have earlier assembly time. For comparison, we
sub-divide the halos into three narrow $t_1$ bins, and show the
corresponding $z_{\rm f}$ - $1+\delta(6\mpc)$ relation in the same
panel of Fig. \ref{fig_zfe}. As one can see, at a fixed value of
$t_1$ the correlation between $z_{\rm f}$ and $1+\delta(6\mpc)$ is
almost absent. In contrast, for fixed local overdensity, the
dependence of $z_{\rm f}$ on $t_1$ is almost as strong as the
overall trend (the upper-right panel of Fig. \ref{fig_zfe}). In
particular, the correlation is significant even in underdense
regions (see the blue-dash dot line in the upper-right panel of
Fig. \ref{fig_zfe}). As a further test we show, in the bottom
panel of Fig. \ref{fig_zfe}, the correlation between $z_{\rm f}$
and $t_1$ separately for halos in clusters, filaments, sheets and
voids, as defined in Appendix B. Clearly, the dependence of
$z_{\rm f}$ on $t_1$ is not affected significantly by the
morphology of the environment. We have made calculations using
overdensities in spheres with radii other than $6\mpc$ and found
that our results are robust against such change. The results are
similar for halos with $\log (M_h/\msun)<11.7$. However, for halos
with $\log (M_h/\msun)>11.7$, the environmental effect is too
weak, compared to the correlations among different environmental
indicators, to break the degeneracy.

All these results unequivocally demonstrate that the amplitude of
the halo tidal field, represented by $t_1$, is the primary
environmental factor that has the most important impact on halo
assembly. This is consistent with suggestion made by Wang et al.
(2007a) that the large-scale tidal field may accelerate mass
around halos, especially low-mass ones, and truncate their mass
accretion. The dependence of halo assembly time on local
overdensity and on the morphology of the environment is the
secondary effect induced by the tidal force. As shown in Fig.
\ref{fig_zfe}, the tidal effect exists even for $t_1<0.005$, where
the self-gravity is much larger than the tidal force. Thus, it is
tidal truncation rather than tidal stripping that is responsible
for the halo assembly bias.

\subsection{Mass fraction in substructures}\label{sec_sub}

 Fig.\,\ref{fig_t1sub} shows the correlations between
the tidal parameter, $t_1$, and the mass fraction in total
substructure, $f_{\rm s}$, in the bound substructure, $f_{\rm
bs}$, and in the unbound substructure, $f_{\rm us}$. Here SUBFIND
(Springel et al. 2001) was used to identify subhalos; bound
substructures are defined to be subhalos with negative total
energy relative to the main halo, while unbound substructures are
unbound subhalos plus fuzz. Clearly, the amounts of total
substructures tend to be larger in high-$t_1$ environments. Thus,
halo clustering strength is expected to depend on the substructure
fraction (e.g. Gao \& White 2007; Ishiyama, Fukushige, \& Makino
2008). However, the environmental dependence of the substructure
fraction becomes very different after the removal of the unbound
component; the dependence of bound component on $t_1$ is totally
absent for massive halos, and there is a weak, but significant,
trend that $f_{\rm bs}$ actually declines with increasing $t_1$
for low-mass halos. Such halo mass dependence of the $f_{\rm
bs}$-$t_1$ relation is very similar to the halo mass dependence of
the $z_f$-$t_1$ relation. Unbound substructure fraction also
increases with increasing $t_1$. In particular, this correlation
is much stronger and is almost halo mass independent. These
results suggest that the $t_1$-$f_{\rm s}$ correlation is
dominated by the environmental dependence of $f_{\rm us}$. Since
$f_s$ is a measure of the mass a halo assembles after the
formation of its main halo and $f_{\rm us}$ is the unbound
fraction, the results reflect the fact that regions of stronger
tidal field, which are also of higher densities, provide more
material for halo to accrete, but the tidal effect in these
regions are also stronger so that $f_{\rm us}$ increases rapidly
with $t_1$.

Similar results are obtained when local overdensity or halo bias
are used instead of the tidal field as the environmental
indicator. In order to examine which environmental property is
more responsible to the strong dependence of $f_{\rm us}$ on
environment, we show  $f_{\rm us}$ as a function of
$1+\delta(6\mpc)$ in the upper-left panel of Fig. \ref{fig_fuse},
together with the results for halos residing in environments with
the same $t_1$. In comparison, we also show $f_{\rm us}$ as a
function of $t_1$ at fixed $\delta(6\mpc)$ and for given types of
large-scale structure. The results clearly demonstrate that it is
the tidal field that plays the dominating role in affecting the
unbound fraction. Note that the figure only presents the results
for halos with $12<\log (M_h/\msun) <12.5$ and using overdensity
on a scale of $6\mpc$, but our tests using overdensity on
different scales and halos with $12.5<\log (M_h/\msun)<13$ and
$13<\log(M_h/\msun)<13.5$ lead to the same conclusion.

According to our definition, only when $t_1\gs 1$ can the tidal
force overcome halo's self-gravity to cause significant stripping.
However, the dependence on $t_1$ extends all the way to
$t_1<0.005$, and is significant even in underdense regions and in
sheet-like structure (Fig. \ref{fig_fuse}). This suggests that
this effect cannot be produced via tidal stripping. Alternatively,
large-scale tidal field may accelerate the material around a halo,
causing them to move quickly relative to the halo (e.g. Wang et
al. 2007a; Hahn et al. 2009; Fakhouri \& Ma 2009). Some of these
energetic particles and satellites may be falling into dark matter
halos but may not be bound to them. One support for this
hypothesis is the existence of a population of ejected halos,
which were once contained in massive halos but eventually would
leave their hosts (e.g. Wang et al. 2009b).

As mentioned above, there are two environmental effects that can
affect a halo's assembly. On the one hand, the amount of the
material that fuels the accretion increases with local density.
Halos in high-density regions are thus expected to have higher
$f_{\rm s}$ and lower $z_{\rm f}$. On the other hand, the tidal
field in a high-density region is on average stronger so that a
larger fraction of the accreted material may become unbound to the
halo, in particular for low-mass halos where tidal force is more
important relative to halo self-gravity. This would make halos in
high-density regions more difficult to grow. The growth of a halo
is therefore the result of the competition between these two
processes, and they affect halo properties in different ways,
depending on the halo mass. For low-mass halos where tidal
suppression of growth is more important, halos in high tidal
force/density regions are expected to have lower $f_{\rm bs}$ and
to form earlier (with higher $z_{\rm f}$). For massive halos, on
the other hand, the two effects may play comparable roles, so that
the environmental dependence of bound substructure and of halo
assembly time is reduced. The two processes acting together
produce a much stronger environmental dependence for $f_{\rm us}$
than for $f_{\rm s}$ regardless of halo masses.

The second process results in a change of the ratio between bound
and unbound components with tidal force \emph{for both low-mass
and massive halos}. To verify this, we make further analysis by
dividing halos into three equally-sized subsamples according to
their $f_{\rm us}$ and examining the $t_1$ - $f_{\rm bs}$
correlation separately for these subsamples. The results for these
four mass bins are shown in Fig. \ref{fig_tfbsc}. Significant
anti-correlation between the tidal force and bound substructure
fraction is indeed found for all these subsamples, even though it
is absent in the whole sample of massive halos. Fig.\ref{fig_tzfc}
show the correlation between $t_1$ and $z_{\rm f}$ for these
subsamples. Clearly, halos form earlier in high tidal-force
environment for all these subsamples, even though this effect does
not show up in the total sample of massive halos. The amount of
material with energy sufficiently low to be accreted is smaller in
an environment of stronger tidal field, so that the amount of
bound substructure and halo growth are both suppressed. Our
finding naturally explains the apparent conflict in the
bias-$z_{\rm f}$ relation, the bias - $f_{\rm s}$ relation and the
$z_{\rm f}$ - $f_{\rm s}$ relation (see Gao \& White 2007). Note
that the absence of strong $t_1$-$f_{\rm bs}$ and $t_1$-$z_{\rm
f}$ correlations for the lowest $f_{\rm us}$ subsample is due to
the use of non-zero $f_{\rm us}$-bin size. Our test showed that
the correlations indeed appear when this subsmaple is divided
further into narrower $f_{\rm us}$ bins.

\subsection{Halo shape and orientation}

\begin{table}
\begin{center}
\caption{The mean cosine of the angle between the principal axes
         of a halo and its orientation} \label{tab_hn}
\begin{tabular}{lccc}
  \hline\hline
  $\log M_h (\msun)$ & $<|\hat{I}_1\cdot\vec{t}_1|>$ & $<|\hat{I}_2\cdot\vec{t}_2|>$ & $<|\hat{I}_3\cdot\vec{t}_3|>$ \\
  \hline
  $[14.5, 15]$& 0.75$\pm$0.02 & 0.60$\pm$0.02 & 0.74$\pm$0.02  \\
  $[14, 14.5]$& 0.732$\pm$0.006 & 0.560$\pm$0.007 & 0.674$\pm$0.007  \\
  $[13.5, 14]$ & 0.698$\pm$0.003 & 0.547$\pm$0.004 & 0.642$\pm$0.004 \\
  $[13, 13.5]$ & 0.647$\pm$0.002 & 0.526$\pm$0.002 & 0.601$\pm$0.002  \\
  $[12.5, 13]$ & 0.600$\pm$0.003 & 0.515$\pm$0.003 & 0.583$\pm$0.003  \\
  $[12, 12.5]$ & 0.567$\pm$0.002 & 0.506$\pm$0.002 & 0.567$\pm$0.002  \\
  \hline
\end{tabular}
\end{center}
\end{table}

Hahn et al. (2007b) found halo major axes are strongly aligned
with that of tidal field. We confirm their results using halo
tidal field. The mean values of $|\hat{I}_1\cdot \vec{t}_1|$ are
listed in Table \ref{tab_hn}. For comparison, we also list the
corresponding results for the intermediate and minor axes.
Clearly, there is a strong alignment between $\hat{I}_1$ and
$\vec{t}_1$, and between $\hat{I}_3$ and $\vec{t}_3$, but the
alignment between $\hat{I}_2$ and $\vec{t}_2$ is much weaker. The
alignments are also stronger for more massive halos.

Since substructures tends to fall into the host halos along the
filament (e.g. Wang, et al. 2005; Altay, Colberg \& Croft 2006),
one might think that the alignments are dominated by the presence
of substructure. In order to test this, we make calculations using
only particles contained by the main halos to calculate the tensor
of inertia. There is little change in the alignments,
demonstrating that the main halos are also aligned with the tidal
field.

In the upper panels of Fig. \ref{fig_tsjht}, we show the average
of $|\hat{I}_k\cdot \vec{t}_k| (k=1, 2, 3)$ as a function of
$t_s$. The alignment of major axes shows significant dependence on
$t_s$; halos in regions with higher values of $t_s$ (i.e. where
the tidal field is more anisotropic) tend to be more strongly
aligned with the tidal field. This trend is either weak or absent
for the intermediate and minor axes.

These alignments may be the primary reasons for some of the
alignments observed in simulations and observations, including the
alignment between the orientations of neighboring galaxy clusters
(Binggeli 1982; Chambers et al. 2002), the alignment between the
orientation of the brightest cluster/group galaxy and the
distribution of its satellites (Carter \& Metcalfe 1980; Wang et
al. 2008b; Faltenbacher et al. 2007; Kang et al. 2007), the
alignment between the galaxy/galaxies cluster orientations and
large-scale structure (Hirata et al. 2007; Faltenbacher et al.
2009; Wang et al. 2009c; Okumura \& Jing 2009; Okumura, Jing \& Li
2009), and furthermore the dependence of the alignments on halo
mass (e.g. Jing 2002; Yang et al. 2006b). One possibility is that
the accretion onto a dark matter halo occurs through a dominating
filament, so that the halo is elongated along the filament (e.g.
Van Haarlem \& Van deWeygaert 1993; Altay et al. 2006). Since the
major axis of the tidal field is expected to trace well the
direction of the local filamentary structure (Hahn et al. 2007b;
Zhang et al. 2009), a strong alignment between $\hat{I}_1$ and
$\vec{t}_1$ can be produced. However, such a mechanism is
difficult to explain the $\hat{I}_3$ - $\vec{t}_3$ alignment,
which is as strong as the $\hat{I}_1$ - $\vec{t}_1$ alignment.
Alternatively and perhaps more likely, the collapse of a density
perturbation to form a halo may be affected by the tidal field, as
in the ellipsoidal collapse model (see Sheth, Mo \& Tormen 2001;
Shen et al. 2006), and the halo orientation is a result of the
corresponding triaxial collapse.

In addition to the alignment of halos with tidal field, we also
study how the axis ratio of a halo, for example $I_3/I_1$, is
correlated with the tidal field. In Fig. \ref{fig_th31c}, we show
the median value of $I_3/I_1$ as a function of $t_1$ ( black solid
lines) for halos of four mass bins. Only most massive sample shows
some weak trend that halos in stronger tidal field tend to be more
spherical. Moreover, the axis ratio is independent of $t_s$ for
halos of all masses. Bett et al. (2007) investigated the
environmental dependence of halo shape and found that spherical
halos are more strongly clustered than the more aspherical ones.
This is different from our results. To compare with their results
more directly, we compute the halo bias as a function of $I_3/I_1$
(see Appendix B for how halo bias is defined and computed).
Similar to the $t_1$ - $I_3/I_1$ correlation, no significant trend
as seen by Bett et al. is found. The difference may be due to the
fact that Bett et al. only considered halos in quasi-equilibrium
state. Since the virialization of a halo is related to the
substructure fraction (Shaw et al. 2006), we divide the halos into
three equally-sized subsamples according to $f_{\rm us}$ and
re-examine the environmental dependence separately for these
subsets of halos. The results are shown in Fig.\ref{fig_th31c} to
compare with the total sample. In each subsmaple, halos are more
spherical in high tidal force region. As shown in
Fig.\ref{fig_fslh}, the axis ratio depends strongly on the bound
substructure fraction, $f_{\rm bs}$. As $t_1$ increases, $f_{\rm
bs}$ decreases and the halo becomes more spherical. This effect is
absent for the full sample because higher $f_{\rm us}$ halos are
more elongated (Fig. \ref{fig_fslh}) and have a stronger tendency
to reside in higher tidal force region (Fig. \ref{fig_t1sub}),
compensating the trend.

\subsection{Halo spin}

According to the tidal torque theory, the angular momenta of dark
matter halos is generated by large-scale tidal field (e.g. Peebles
1969; Doroshkevich 1970).  In the left panel of Fig.
\ref{fig_t1ol} we show the dependence of $\lambda$ on $t_1$.
Clearly halos tend to spin faster in stronger tidal field, and
this trend is stronger for massive halos. However the dependence
is rather weak. The value of $\lambda$ increases by a factor less
than two as the value of $t_1$ increases by an order of magnitude,
much weaker than the linear dependence expected from the tidal
torque theory. Note that the dependence on $t_s$ is absent. It is
known that halo clustering strength increases with halo spin (e.g.
Bett et al. 2007). In order to make a direct comparison between
these two environmental properties, we present $\lambda$ as a
function of $\delta(6\mpc)$, instead of halo bias, versus
$\lambda$ in the right panel of Fig. \ref{fig_t1ol}. The
correlation strength is similar to that with tidal
force\footnote{We have also tested using overdensity on other
scales instead of 6 $\mpc$ and obtained very similar results}.

\begin{table}
\begin{center}
\caption{The mean cosine of the angle between spin axis of halo
and the principle axes of halo tidal field} \label{tab_gn}
\begin{tabular}{lccc}
  \hline\hline
  $\log M_h (\msun)$ & $<|\hat{j}\cdot\vec{t}_1|>$ & $<|\hat{j}\cdot\vec{t}_2|>$ &
$<|\hat{j}\cdot\vec{t}_3|>$ \\
  \hline
  $[14.5, 15]$& 0.39$\pm$0.02  &  0.59$\pm$0.02 &  0.49$\pm$0.02  \\
  $[14, 14.5]$& 0.443$\pm$0.007 & 0.550$\pm$0.007 & 0.495$\pm$0.007  \\
  $[13.5, 14]$ & 0.452$\pm$0.003 & 0.541$\pm$0.003 & 0.491$\pm$0.004 \\
  $[13, 13.5]$ & 0.472$\pm$0.002 & 0.532$\pm$0.002 & 0.484$\pm$0.002  \\
  $[12.5, 13]$ & 0.489$\pm$0.003 & 0.522$\pm$0.003 & 0.480$\pm$0.003  \\
  $[12, 12.5]$ & 0.503$\pm$0.002 & 0.510$\pm$0.002 & 0.479$\pm$0.002  \\
  \hline
\end{tabular}
\end{center}
\end{table}

Since the frequency of halo merging increases with local density
(e.g. Fakhouri \& Ma 2009), the above correlations can also be
interpreted via the merger scenario (Gardner 2001; Maller et al.
2002). To gain more insight, we examine the alignment between the
halo angular momentum vector, $\hat{j}$, and the three principal
axes of the tidal field, $\vec{t}_k$. The mean values of the dot
product $|\hat{j}\cdot\vec{t}_k|$ for various halo masses are
listed in Table \ref{tab_gn}. As one can see, halos obviously tend
to spin around axes perpendicular to the major axes of the
large-scale tidal field. The strength of the alignment decreases
with decreasing halo mass and becomes absent for halos of
$\log(M_h/\msun)\simeq 12$. Some recent studies have found a weak
but significant alignment between the spin of low-mass halos
($\log(M_h/\msun)\simeq 11$) and the orientation of the
large-scale mass distribution (Arag$\acute{{\rm o}}$n-Calvo et al.
2007; Paz et al. 2008; Zhang et al. 2009).  Using the major axis
of the halo tidal field to approximate the orientation of large
scale mass distribution, we also detect such an alignment signal
for halos in a similar mass range.

There is also a weak tendency for the spin axis to be
perpendicular to the minor axis of tidal field, and the strength
is weaker than that to the major axis of the tidal field in the
mass ranges in consideration. In contrast,  halo spin is {\it
aligned} with the intermediate axis of tidal field (see Table
\ref{tab_gn}). Such an alignment is a natural prediction of the
tidal torque theory (e.g. Porciani et al. 2002; Lee \& Erdogdu
2007), and so our result provides support to the tidal torque
origin for the halo angular momentum. However, except for massive
halos with $14.5<\log(M_h/\msun)<15$, the alignment strength we
find (see Table \ref{tab_gn}) is much weaker than the value $0.59$
predicted by the tidal torque theory (Porciani et al.
2002)\footnote{Note that the major (minor) axis of the tidal field
defined in Porciani et al. (2002) correspond to the minor (major)
axis defined here.}, particularly for low-mass halos. Since the
tidal torque theory is based on linear density field, the
discrepancy may be due to non-linear evolution. Indeed, using
N-body simulations, Porciani et al. (2002) did not find any
alignment between halo spins and initial tidal field and concluded
that non-linear effects completely erase the correlation. However,
Porciani et al. (2002) focused on relatively low-mass halos, for
which the alignment strength is weak according to our results.
Significant correlation does exist for massive halos.

It has been suggested that the orientations of halo spin vectors
are correlated with the morphology of the nearby large-scale mass
distribution (e.g. Arag$\acute{{\rm o}}$n-Calvo et al. 2007; Hahn
et al. 2007a; Zhang et al. 2009). For instance, the spin axes of
halos in sheets tend to lie in the sheet, while halos in filaments
have a tendency to spin around axes perpendicular to the
filaments. Similar trends have also been claimed in observational
data. Navarro, Abadi \& Steinmetz (2004) found that the spin axes
of nearby disk galaxies tend to lie on the supergalactic plane.
Trujillo, Carretero \& Patiri (2006) analyzed a large sample of
spiral galaxies and found that galaxies that are located on the
surfaces of cosmic voids have spin axes tending to be parallel to
the surfaces.

As discussed above, the $t_s$ parameter defined in our analysis
can be used to quantify the morphology of nearby mass/halo
distribution. In the lower three panels of Fig. \ref{fig_tsjht} we
show the mean value of $|\hat{j}\cdot\vec{t}_k|$ as a function of
$t_s$. As one can see, the alignment of spin with the intermediate
axis of the tidal field is almost independent of $t_s$, while the
other two alignments show clear trend with $t_s$. Halos in low
$t_s$ environment tend to have their spins parallel to the major
axes and perpendicular to the minor axes of tidal field, while
halos in high $t_s$ environment tend to spin around axes
perpendicular to the major axes and parallel to the minor axes of
tidal field. Here again, the trend is stronger for more massive
halos. These trends are in broad agreement with previous results
based on the morphology of the large-scale mass distribution.

We have also searched for possible correlation between $t_1$ and
the strengths of the spin - tidal field alignments, and found that
any such correlation is either absent or very weak. This suggests
that the dependence of the spin alignment on the morphology of the
large-scale mass distribution is due to the difference in the
`shape', not in the magnitude, of the tidal fields in different
environments. According to the tidal torque theory, the alignment
between the spin axis and the tidal directions at redshift zero is
related to the correlation between the tidal field and inertia
tensor of proto-haloes in initial condition (Porciani et al. 2002
). It is possible that the $t_s$-dependence may be due to the fact
that the initial correlation varies with the shape of tidal field,
instead of its strength.

The results presented above show clearly that halo spins are
related to the tidal torques produced by the large-scale mass
distribution. However, the correlation between halo spins and
tidal field is much weaker than that predicted by the simple tidal
torque theory, particularly for low-mass halos, suggesting that
the relationship between halo spins and tidal fields is
complicated. Tidal field can not only exert torques on halos, but
also affect halo assembly histories. As we have seen in Subsection
\ref{sec_at}, halos residing in stronger tidal fields on average
have higher assembly redshifts, particularly for low-mass halos,
and such trend may be a result of tidal truncation of mass
accretion by halos. As shown in Fig. \ref{fig_zfhp}, there is a
clear tendency that halos of the same mass with higher assembly
redshifts spin slower, presumably because the mass that is
prevented from being accreted by the tidal field has on average
higher specific angular momentum. It is thus likely that halo spin
is the result of two competing effects of tidal field: tidal
torque and tidal truncation. For more massive halos where tidal
truncation is less important (see also Fakhouri \& Ma 2009 for
similar results), the correlation between halo spin and tidal
field is stronger, as is seen in our results. For low-mass halos,
on the other hand, the tidal truncation is so important that the
torque effect is significantly reduced.

\section{Summary and discussion}\label{sec_con}

In this paper, we use seven high-resolution $N$-body simulations
to study the correlation between different halo properties, and
between halo properties and large-scale environment. We focus on
the following halo properties: assembly time, mass fraction in
substructure bound and unbound to the main halo, halo spin and
shape. The large-scale tidal field estimated from halos above a
mass threshold is used as our primary quantity to describe the
environment in which a halo resides.

We first examine the relationship between halo properties and find
that most of them are correlated. Young halos tend to have more
substructures, spin more rapidly and be less spherical than their
old counterparts. Halos containing large amount of substructures
generally have higher spin parameter and appear more aspherical.
Halo spin decreases with increasing axis ratio of a halo (i.e. as
halo becomes rounder). All of these correlations are connected,
but the underlying causal processes may be convolved. One possible
process often discussed is the accretion of nearby halos,
especially major mergers. However, mass redistribution, in
particular the ejection of subhalos and mass, may also be
important in producing, at least part of, these correlations.

We then investigate the environmental dependence of halo
properties. Low-mass halos tend to be older in stronger tidal
field/overdensity region. Such dependence is absent for massive
halos. The total substructure fraction, which is higher for
younger halos, has an opposite environmental dependence in the
sense that the substructure fraction increases with the strength
of the tidal field for both low-mass and massive halos. To
understand this discrepancy, we separate the substructures into
bound and unbound components, and find that the unbound component
has a similar but much stronger and closer trend with environment
than the total substructure fraction. The dependence of bound
component on environment differs from that of the unbound
component, but is similar to that of the assembly time: bound
component decreases with increasing tidal force for low-mass halos
and the trend is absent for massive ones. For massive halos of
given unbound substructure, however, both the bound fraction and
the assembly time exhibit strong and significant trends with
environment, similar to the trends seen for low-mass halos.

To gain more insight into these environmental dependencies, it is
important to sort out which environmental effect has the closest
connection to halo properties. Our results demonstrate that the
correlations of assembly time and unbound substructure fraction
with the local overdensity and the morphology of the large-scale
structure are actually induced by the correlations with the
large-scale tidal field. So it is the tidal field that is more
fundamental in driving the environmental dependence detected in
N-body simulations. As suggested by Wang et al. (2007a), tidal
field can accelerate the material around a halo, increasing the
fraction unbound component and suppressing mass accretion into the
halo.

How are these environmental effects, part of which seem to be in
contradiction, produced? Based on our results, we suggest that
environmental effects can act in two different ways. First, the
amount of material that fuels the accretion into halos increases
with local density so that halos in high-density regions are
expected to have higher substructure fraction. Second, the
fraction of accreted material which is unbound to the halo
increases with tidal field because of the acceleration of tidal
field. These two processes combined yield a strong correlation
between the unbound fraction and the strength of tidal field.
However the bound fraction and assembly time of a halo is the
result of the competition between these two processes. For
low-mass halos where the second process is more important, halos
in high tidal force regions tend to have lower bound fraction and
to form earlier. For massive halos, these two effects are
comparable, so that the environmental effect is reduced. A
consequence of the second process is that the ratio of the bound
to unbound fraction should decrease with increasing tidal strength
for both low-mass and massive halos, which is shown clearly in our
results.

We find that halo spin has a mild but significant correlation with
tidal field. Halos have a tendency to spin more rapidly in
stronger tidal field. This suggests that the spin of halos
originates from tidal torque rather than from random mergers.
Using halo tidal field, we find that halo spin vectors tend to lie
perpendicular to both major and minor axes and parallel to the
intermediate axes of tidal field. The alignment with the
intermediate axes of the tidal field is a natural prediction of
the tidal torque theory, and so provides support to the theory. We
also find that these alignments, except that with the intermediate
axis, vary with the `shape' of the tidal field. These findings
provide valuable constraints on the tidal torque theory. However,
the tidal torque effect of a strong tidal field can be compensated
by tidal truncation and weakened by the continuous virialization
process.

In addition to the strong alignment between the major axes of
halos and large scale tidal field, we also find a strong minor
axes alignment. These alignments may be the primary origin of
other alignments detected in simulations and observations. For
example, the alignment between the orientations of neighboring
galaxy clusters, between the orientation of the brightest
cluster/group galaxies and the distribution of their satellites,
and between the orientations of galaxies and the large-scale
structure. The strength of the minor-axis alignment is as strong
as that of the major axis and can not be produced via the infall
of material along filament. More likely, the halo orientation may
be a result of triaxial collapse modulated by tidal field.
Finally, we examine the relationship between halo axis ratio and
environment, and find no strong trend. Nevertheless, for halos of
given unbound substructure fraction, halos become more spherical
in stronger tidal field.

Based on this study, we find that using halo tidal field has a
number of advantages: (i) It accurately represents the large-scale
tidal field, while the tidal field calculated from total mass
distribution is affected by the choice of smoothing scale and
contaminated by the internal tidal field produced by halo's
self-gravity; (ii) The shape parameter, $t_s$, of the halo tidal
field provides a continuous quantity describing the distribution
of the surrounding mass distribution, and hence can be used to
quantify the dependence of halo properties on the morphology of
the large-scale structure; (iii) This method can be directly
applied to observational data, especially to group catalogs where
halo mass information is available (e.g. Yang et al. 2005, 2007);
(iv) Among all the environmental indicators considered here, the
magnitude of the halo tidal field, $t_1$, has the strongest
correlation with halo properties.

It is worthwhile to point out that the correlations among assembly
time, bound substructure fraction and axis ratio, and between
unbound substructure fraction and spin parameter, are stronger
than the correlations between the quantities across these two
parameter sets. These two sets of quantities are also different in
their environmental dependencies. In regions of strong tidal
field/high local density, halos tend to form \emph{earlier},
contain \emph{less} bound substructure and be \emph{more
spherical}, \emph{but} have a tendency to contain \emph{more}
unbound substructure and spin \emph{faster}. This suggests that
these two sets of quantities may have different origins.

It has been claimed that the environmental dependencies of the
assembly time and total substructure fraction are inconsistent
with the correlation between the assembly time and total
substructure fraction (e.g. Gao \& White 2007). The environmental
dependencies of axis ratio and spin parameter also seem to be in
conflict with the correlation between them. Our results show that
these apparent discrepancies can be understood, because
environmental effects can act in two different ways, as discussed
above.

Our results have important implications for understanding how
galaxy properties are correlated with their environments. In our
current model of galaxy formation, galaxies are assumed to form in
dark matter halos, and so the properties of a galaxy are expected
to be correlated with the properties of its host halo. For
instance, the spin of a disk is expected to be related to the spin
of its host (e.g. Mo et al. 1998), the central galaxies in galaxy
clusters may be aligned with the host halos (e.g. Yang et al.
2006b; Kang et al. 2007), and the stellar population and
morphology of galaxies may be related to the assembly histories of
their host halos. Thus, the dependence of halo properties on the
large scale tidal field found in the present paper would imply
correlations of these galaxy properties with large-scale tidal
field. Since the halo tidal field can be estimated from
observation, these correlations can all be tested observationally.
Furthermore the large-scale tidal field may also affect galaxy
formation directly. For example, the large-scale tidal field is
expected to promote the formation of large-scale pancakes and
filaments. The shocks associated with the formation of such
structures, especially the inside massive clusters,  may heat the
surrounding gas, modulating gas accretion by galaxies from the
intergalactic medium and affecting the properties of galaxies (Mo
et al. 2005; Dolag et al. 2006). Such effects may be studied by
analyzing the correlation between the gas and galaxy
distributions, as well as their correlations with the halo tidal
field.

\section*{Acknowledgment}

We thank Volker Springel for kindly providing his SUBFIND code,
and Ramin Skibba, the referee of the paper, for helpful comments
that greatly improved the presentation of this paper. HYW is
Supported by NSFC 11073017, the Fundamental Research Funds for the
Central Universities and Outstanding Phd Thesis Found of CAS. HJM
would like to acknowledge the support of NSF AST-0908334. This
work is partly supported by NSFC (10821302, 10878001, 10925314),
by the Knowledge Innovation Program of CAS (No. KJCX2-YW- T05), by
973 Program (No. 2007CB815402) and by the CAS/SAFEA International
Partnership Program for Creative Research Teams (KJCX2-YW-T23). YW
is supported by the Fundamental Research Funds for the Central
Universities, NSF-10903020, Research Foundation for Talented
Scholars.

\begin{figure}
\epsscale{1}\plotone{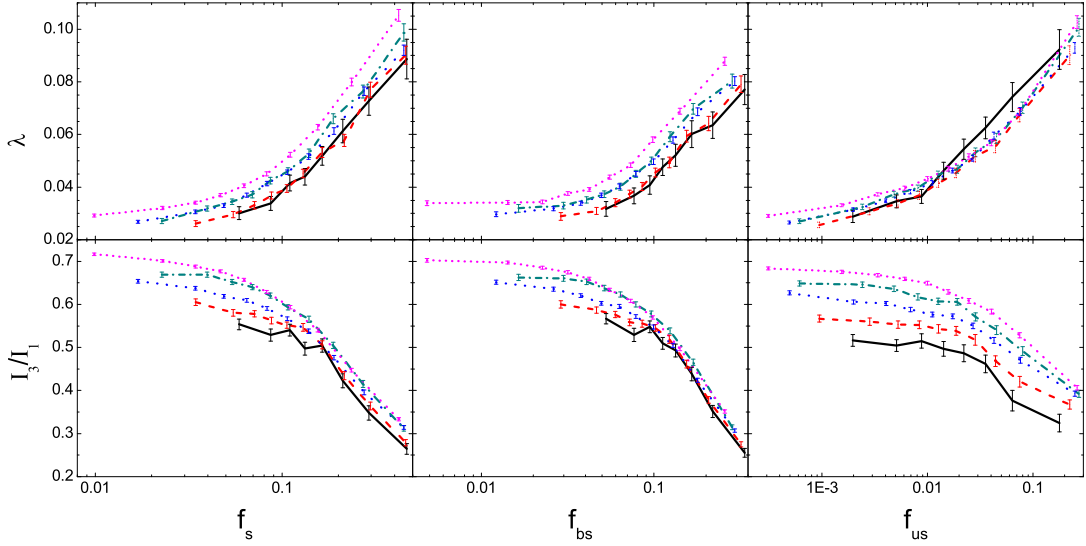}
 \caption{Median values of $\lambda$ and $I_3/I_1$ as functions
of $f_{\rm s}$, $f_{\rm bs}$ and $f_{\rm us}$ for halos in the
same five mass bins as shown in Fig.\,\ref{fig_zfhp}}.
\label{fig_fslh}
\end{figure}

\begin{figure}
\epsscale{1}\plotone{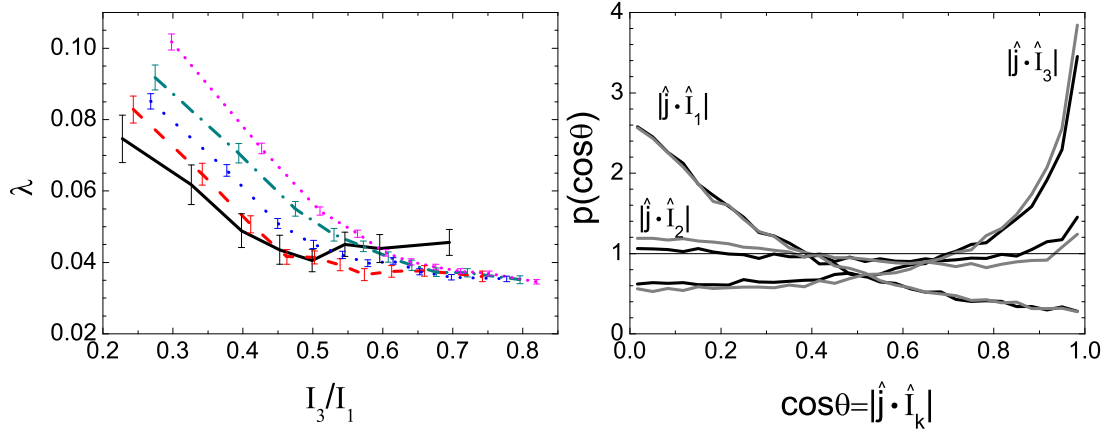}
 \caption{The left panel shows the median $\lambda$ as a function
of short-to-long axis ratio $I_3/I_1$ for halos for halos in the
same five mass bins as shown in Fig.\,\ref{fig_zfhp}. The right
panel shows the probability distribution of the cosine of the
angle between the spin vector, $\hat{j}$, and the three principal
axes $\hat{I}_k$. The black and grey lines are for halos with $13<
log(M_h/\msun)< 14.5$ and
 $12< log(M_h/\msun)< 13$, respectively. The horizontal line
indicates a random distribution.}\label{fig_js}
\end{figure}

\clearpage

\begin{figure}
\epsscale{1}\plotone{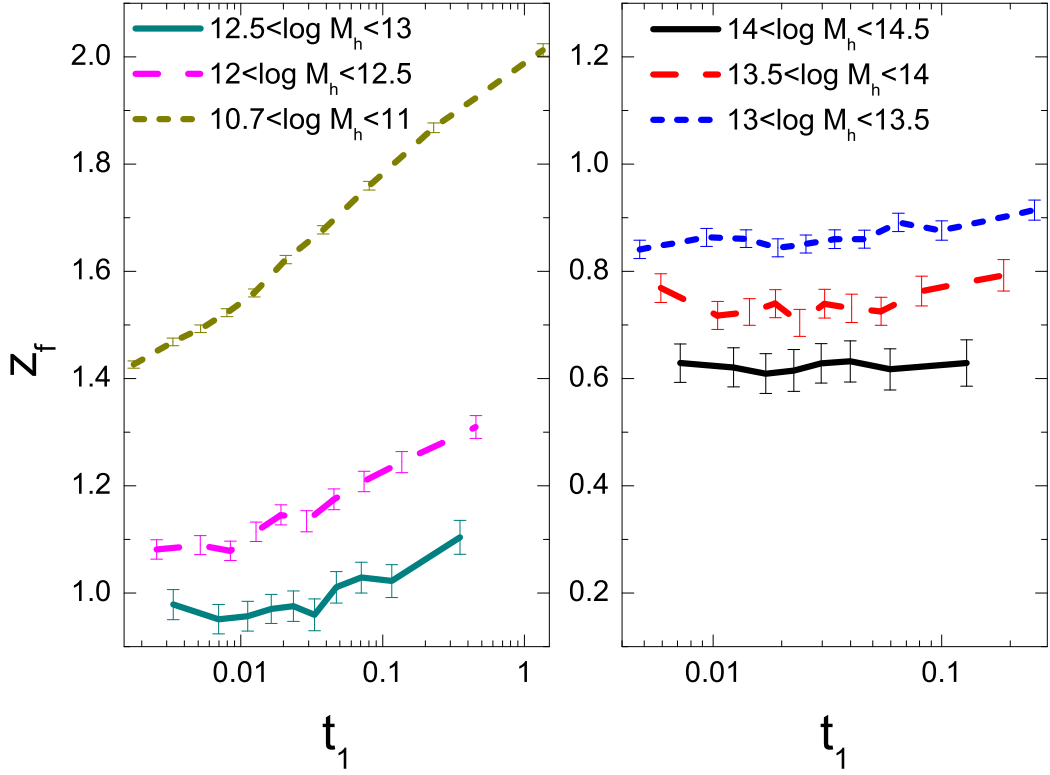} \caption{The median assembly
time, $z_{\rm f}$, as a function of the tidal field $t_1$ for
halos of different masses (in units of $\msun$).} \label{fig_zft}
\end{figure}

\begin{figure}
\epsscale{1}\plotone{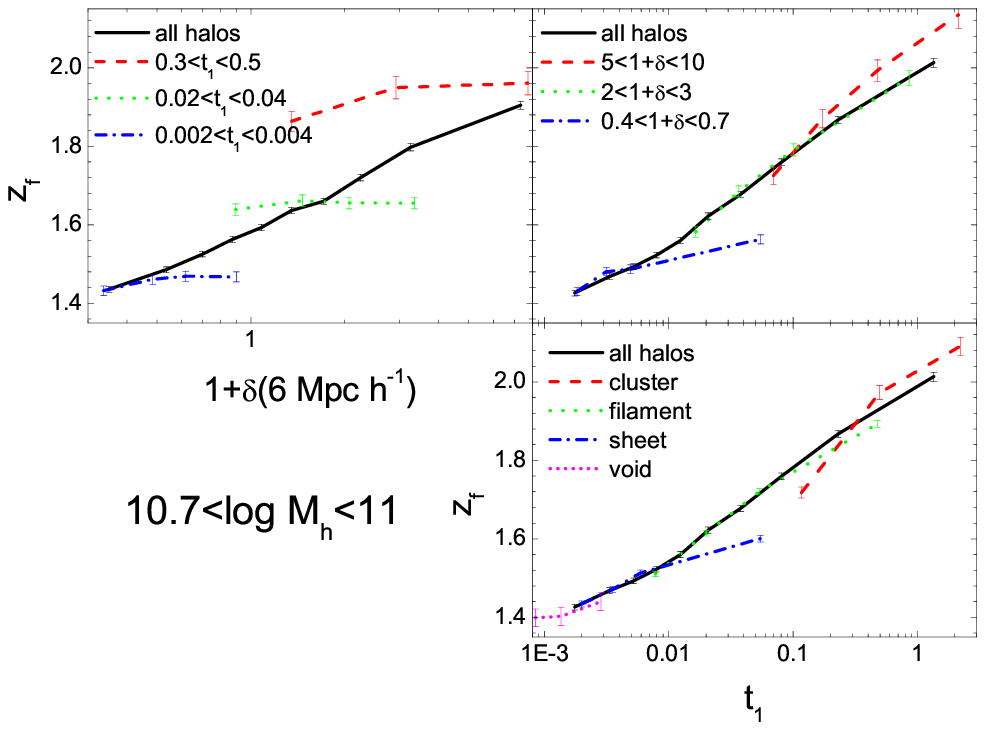} \caption{The black solid line in
each panel shows the median $z_{\rm f}$ as a function of local
overdensity (upper left panel) and $t_1$ (the two right panels)
for halos of $10.7<log(M_h/\msun)<11$. The colored lines show the
results at fixed $t_1$ (upper left panel), fixed local overdensity
(upper right panel), and for given types of large-scale structure
(lower right panel), as indicated in each panel.} \label{fig_zfe}
\end{figure}

\begin{figure}
\epsscale{1}\plotone{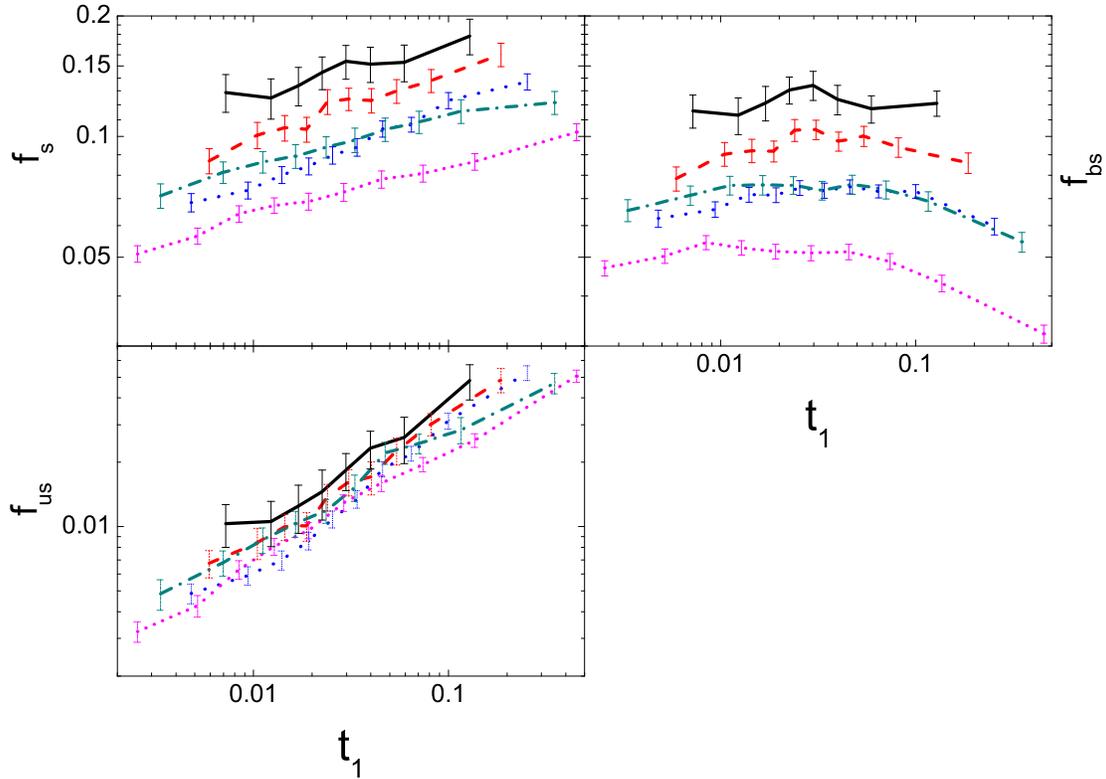} \caption{The median values of
halo properties $f_{\rm s}$, $f_{\rm us}$ and $f_{\rm bs}$ as
functions of $t_1$. Results are shown for halos in the same five
mass bins as in Fig. \ref{fig_zfhp}.} \label{fig_t1sub}
\end{figure}

\begin{figure}
\epsscale{1}\plotone{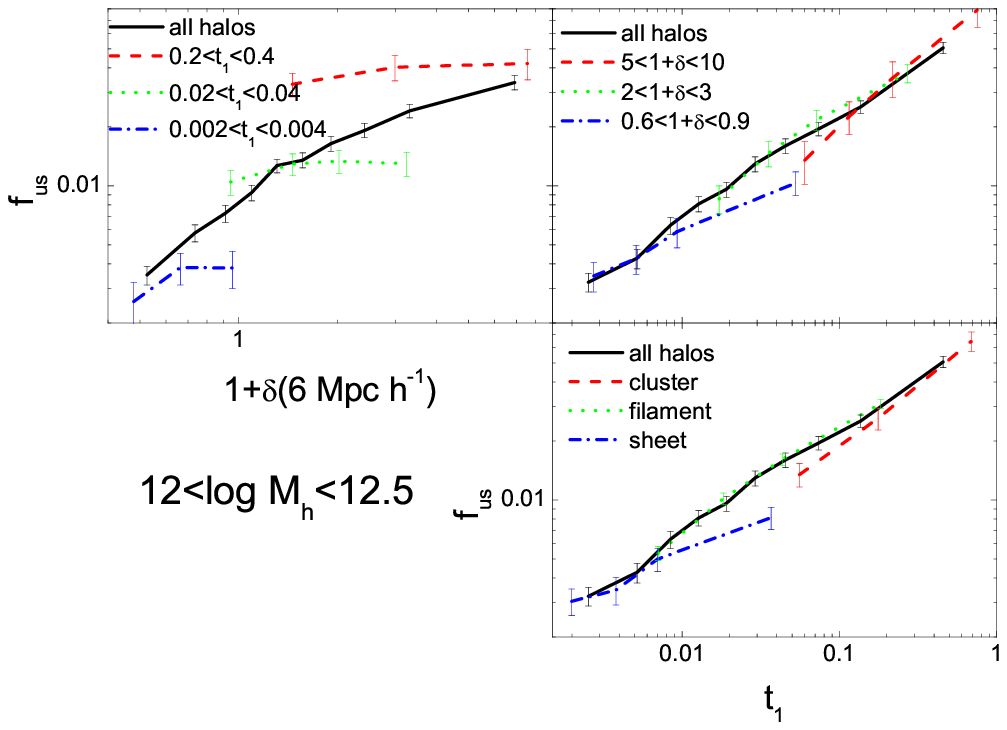} \caption{ The black solid line
in each panel shows the median $f_{\rm us}$ as a function of local
overdensity (upper left panel) and $t_1$ (the two right panels)
for halos of $12<log(M_h/\msun)<12.5$. The colored lines show the
results at fixed $t_1$ (upper left panel), fixed local overdensity
(upper right panel), and for given types of large-scale structure
(lower right panel), as indicated in each panel.} \label{fig_fuse}
\end{figure}

\begin{figure}
\epsscale{1}\plotone{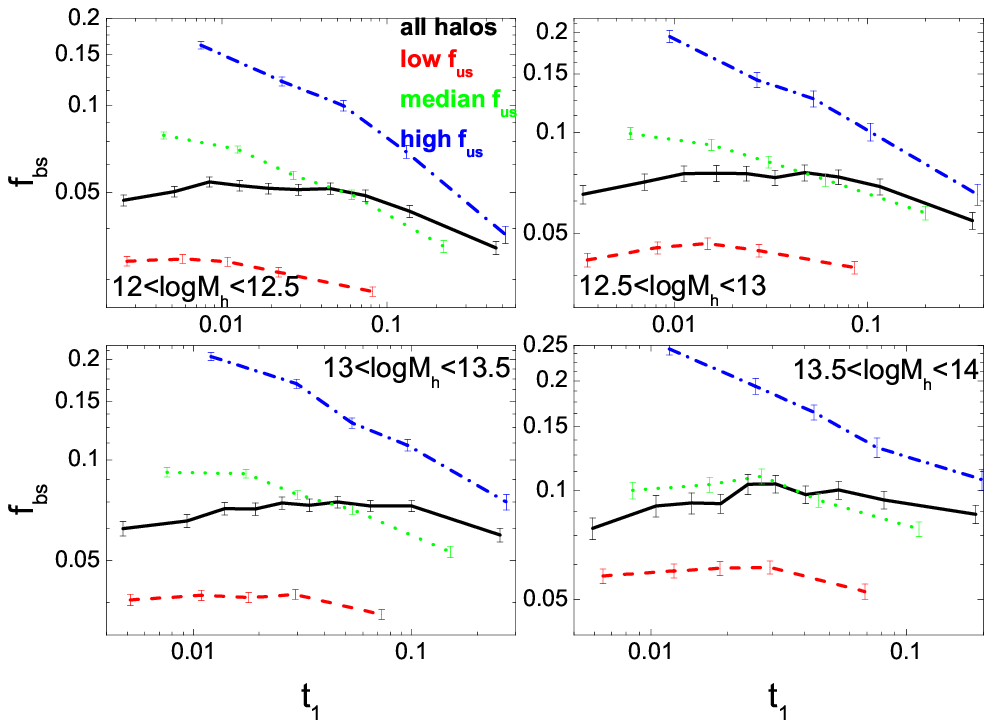} \caption{Here halos of a given
mass are divided into three equal-size subsamples according to
their $f_{\rm us}$. The colored lines show $f_{\rm us}$ as a
function of $t_1$ for these three subsmaples, while the black line
shows that for the total sample.} \label{fig_tfbsc}
\end{figure}

\begin{figure}
\epsscale{1}\plotone{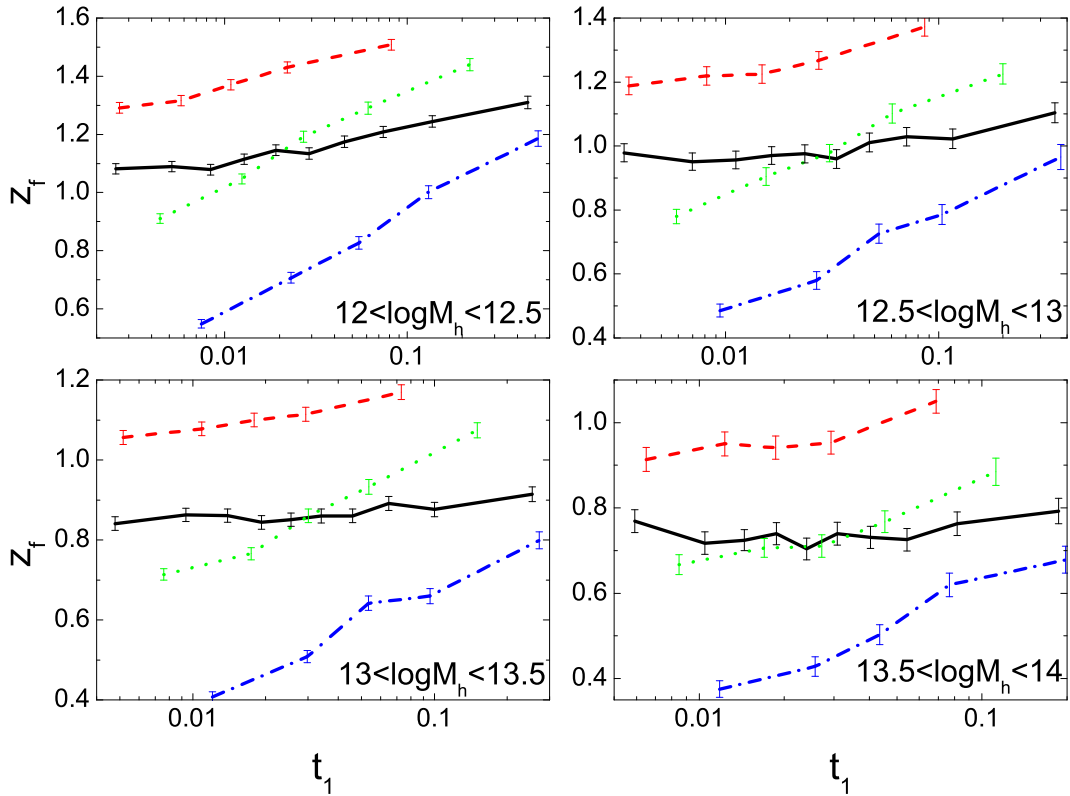} \caption{The same as
Fig.\ref{fig_tfbsc} but for the halo assembly time $z_{\rm f}$ as
a function of $t_1$.} \label{fig_tzfc}
\end{figure}

\begin{figure}
\epsscale{1}\plotone{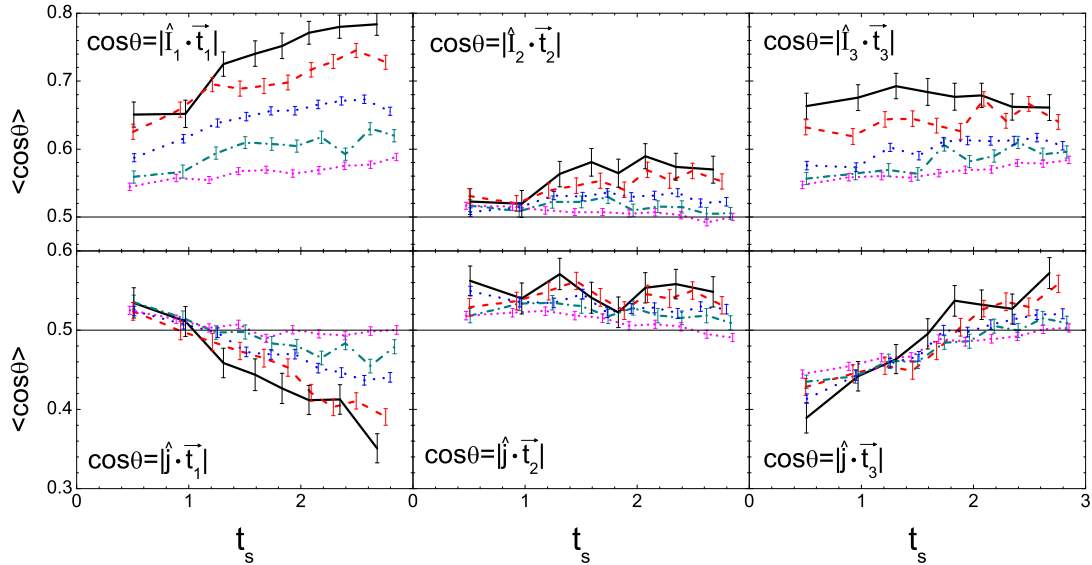} \caption{The upper three
panels show the mean of $|\hat{I}_k\cdot\vec{t}_k|$ as a function
of the shape of tidal field, $t_s$. The lower panels show the mean
of $|\hat{j}\cdot\vec{t}_k|$ as a function of $t_s$. Results are
shown for halos in the same five mass bins as in Fig.
\ref{fig_zfhp}. The horizontal lines indicate random
distribution.} \label{fig_tsjht}
\end{figure}

\begin{figure}
\epsscale{1}\plotone{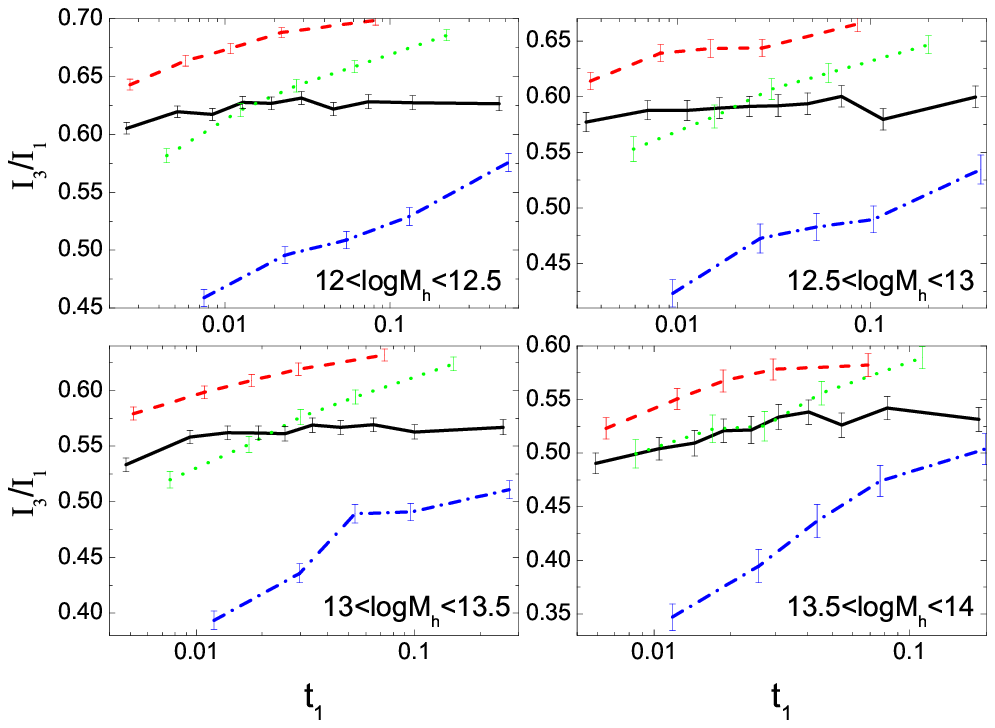}
 \caption{Here halos of a given mass are divided into three
equal-size subsamples according to their $f_{\rm us}$. The colored
lines show $I_3/I_1$ as a function of $t_1$ for these three
subsmaples, while the black line shows that for the total sample.}
\label{fig_th31c}
\end{figure}

\begin{figure}
\epsscale{1}\plotone{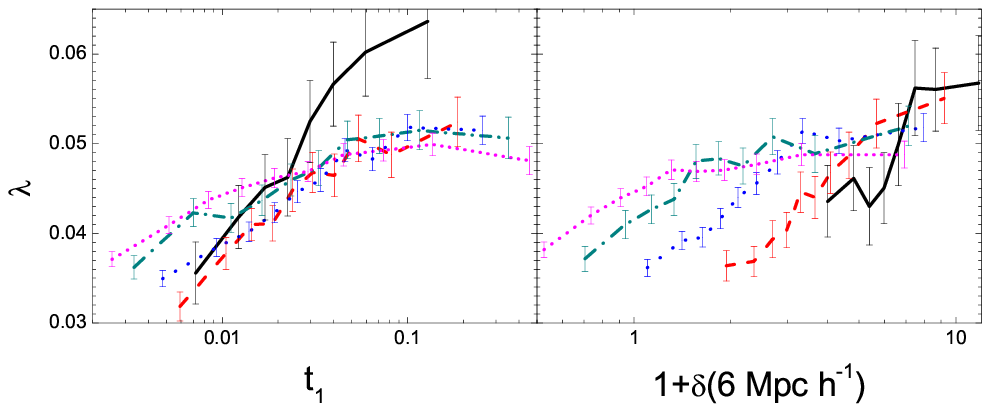} \caption{The median of the spin
parameter, $\lambda$, as a function of tidal force, $t_1$ (left
panel) and local overdensity, $\delta(6\mpc)$ (right panel).
Results are shown for halos in the same five mass bins as in Fig.
\ref{fig_zfhp}.} \label{fig_t1ol}
\end{figure}

\appendix{}

\section{Notes on tidal field from the halo population}

We re-write Eq. (\ref{eq_ft}) as
\begin{equation}
f_t(\vec{t})=
\sum_{i=1}^{N}\frac{R_{i}^3}{2r_i^3}(1+3\cos{2\theta_i})=\sum_{i=1}^{N}\frac{R_i^3}{r_i^3}(3|\vec{t}\cdot\vec{r}_i|^2-1)\,,
\end{equation}
where $\vec{r}_i$ is the unit vector from halo `$i$' to the halo
in question, and the symbol `$\cdot$' means dot product.
$\vec{t}_1$ and $\vec{t}_3$ are defined so that the tidal forces
reach local extrema along these two directions. The necessary
condition for local extrema is that the gradient of the function
$f_t$ is zero along these two directions:
\begin{equation}
{\rm grad}f_t(\vec{t})|_{\vec{t}=\vec{t}_k}=0
\end{equation}
where $k=1,3$. Thus $\vec{t}_k$ satisfy
\begin{equation}
\sum_{i=1}^{N}\frac{R_{i}^3}{r_i^3}(\vec{t}_k\cdot\vec{r}_i)(\vec{t}_k\times\vec{r}_i)=0\label{eq_grd}
\end{equation}
where `$\times$' denotes vector product. Only when the vector
$\vec{t}_k$ is parallel to the vector
$\sum_{i=1}^{N}\frac{R_i^3}{r_i^3}(\vec{t}_k\cdot\vec{r}_i)\vec{r}_i$,
is the left term of this equation equal to 0. So
\begin{equation}
\sum_{i=1}^{N}\frac{R_i^3}{r_i^3}(\vec{t}_k\cdot\vec{r}_i)\vec{r}_i=
\alpha_k\vec{t}_k\label{eq_mat}\,.
\end{equation}
Clearly $\vec{t}_k$ and $\alpha_k$ ($k=1,3$) are the eigenvectors
and corresponding eigenvalues of the matrix
$\sum_{i=1}^{N}\frac{R_i^3}{r_i^3}\vec{r}_i\vec{r}_i$, referred to
as the halo tidal tensor. $\vec{t}_3$ is thus perpendicular to
$\vec{t}_1$. Since $\vec{t}_2$ is perpendicular to both
$\vec{t}_1$ and $\vec{t}_3$, we have
$\vec{t}_2=\vec{t}_1\times\vec{t}_3$. It is easy to prove that
$\vec{t}_2$ is the third eigenvector of the halo tidal tensor.

According to the definition of $t_1$, $t_2$ and $t_3$,
\begin{equation}
t_k=3\sum_{i=1}^{N}\frac{R_i^3}{r_i^3}|\vec{t}_k\cdot\vec{r}_i|^2-\sum_{i=1}^{N}\frac{R_i^3}{r_i^3}=3\alpha_k-\sum_{i=1}^{N}\frac{R_i^3}{r_i^3}\label{eq_tj}
\end{equation}
Combining Eqs. (\ref{eq_tj}) and (\ref{eq_mat}), we obtain
\begin{equation}
t_1+t_2+t_3=3\sum_{i=1}^{N}\frac{R_i^3}{r_i^3}(|\vec{t}_1\cdot\vec{r}_i|^2+|\vec{t}_2\cdot\vec{r}_i|^2+|\vec{t}_3\cdot\vec{r}_i|^2)-3\sum_{i=1}^{N}\frac{R_i^3}{r_i^3}
\end{equation}
Since
$|\vec{t}_1\cdot\vec{r}_i|^2+|\vec{t}_2\cdot\vec{r}_i|^2+|\vec{t}_3\cdot\vec{r}_i|^2=1$,
we have $t_1+t_2+t_3=0$. By definition, $t_1>t_2>t_3$, so that
$t_1>0$ and $t_3<0$.

\section{Notes on other environmental indicators}

\subsection{Local overdensity and bias parameter} \label{sec_ovd}

One of the commonly used environmental parameters of galaxies and
dark matter halos is the local overdensity (e.g. Dressler 1980;
Lemson \& Kauffmann 1999; Maulbetsch et al. 2007). Here we adopt
$\delta (6\mpc)$, the overdensity of dark matter within a sphere
of radius $6\mpc$ around each dark matter halo, as one of our
environmental indicators.

A related measure is the halo bias parameter. For a given set of
halos, it is defined as
\begin{equation}\label{b_definition}
b=\frac{\langle\delta_{\rm hm}(R)\rangle}
  {\langle\delta_{\rm mm}(R)\rangle}\,,
\end{equation}
where  $\langle\delta_{\rm hm}(R)\rangle$ is the average
overdensity of dark matter within a sphere of radius $R$ around
the set of halos in question, and $\langle\delta_{\rm
mm}(R)\rangle$ is the average overdensity within all spheres of
radius $R$ centered on dark matter particles. Note, however, while
quantities like $\delta (6\mpc)$ can be used to indicate the
large-scale environment for a given halo, so that one can study
how halo properties changes with environment, the bias factor $b$
is defined for a population of halos, so that it can be use to
describe the average environment of halos of similar properties.

\subsection{Mass tidal field and the morphology of large scale structure}
\label{sec_tm}

We describe the tidal field of dark matter distribution through
the tidal tensor defined as
\begin{equation}\label{eq_tij}
{\cal T}_{ij}=\partial_i\partial_j\phi\,,
\end{equation}
where $\phi$ is the gravitational potential. In order to compute
${\cal T}_{ij}$, we first use the cloud-in-cell scheme (Hockney \&
Eastwood 1981) to generate the overdensity field on $1024^3$ grid
points from the discrete distribution of the dark matter particles
in the N-body simulations. We then use the Fast Fourier Transform
to obtain the potential field by solving the Poisson equation,
\begin{equation}
\nabla^2\phi=4\pi G\rho_{\rm m}\delta, \label{eq_pe}
\end{equation}
where $G$ is the gravitational constant, $\delta$ is the
overdensity field smoothed with a Gaussian kernel with some
smoothing mass scale (hereafter SMS), and $\rho_m$ is the cosmic
mean density. We apply the derivative operators to calculate the
tidal tensors at the center of mass of each halo in the simulation
and obtain the eigenvectors $\vec{T}_{1}$, $\vec{T}_{2}$, and
$\vec{T}_{3}$, and the corresponding eigenvalues $T_{1}$, $T_{2}$,
and $T_{3}$ ($T_{1}\leq T_{2}\leq T_{3}$). Note that the Poisson
equation requires that $T_{1}+T_{2}+T_{3}= 4\pi G\rho_{\rm
m}\delta$. We refer to the tidal field estimated in this way as
the mass tidal field, to distinguish the tidal field obtained from
the halo population (see the main text).

The number of positive eigenvalues of the mass tidal tensor has
been used to classify the large-scale environment in which a halo
resides (e.g. Hahn et al. 2007a). If all of the three eigenvalues
are positive, the region is defined as a cluster environment.
Similarly, regions with one or two negative eigenvalues are
defined as filaments or sheets, respectively, while regions with
three negative eigenvalues are defined as voids. In this paper, we
will also use the same definition to classify the environments of
dark matter halos. In particular, following Hahn et al.(2007a; see
also Wang et al. 2009a), we choose a fixed SMS,
$2M_{\ast}$,\footnote{$M_\ast$ is the characteristic mass scale at
which the {\it RMS} of the linear density field is equal to
$1.686$, the critical overdensity for spherical collapse, at the
present time. For the present simulations $\log
(M_\ast/\msun)\approx 12.8$.}, in this analysis.

In principle, the same method can also be used to estimate the
large-scale tidal force around a halo. However, since we are
interested in the tidal fields around halos of various masses, we
found that adopting a fixed SMS underestimates/overestimates the
tidal strength around low-mass/massive halos. Because of this, we
choose to adopt an adaptive SMS which is proportional to the mass
of the halo in question. Our tests showed that a SMS between a
half and two times the halo mass gives similar results, and our
results below uses a SMS which is equal to one times the halo
mass. Thus, while we adopt a fixed SMS of $2M_{\ast}$ to define
the type of environment, an adaptive SMS of $M_h$ is adopted to
calculate the large-scale tidal force around individual halos.

\subsection{Comparison with the halo tidal field}
\label{sec_comde}

\begin{figure}
\epsscale{0.5}\plotone{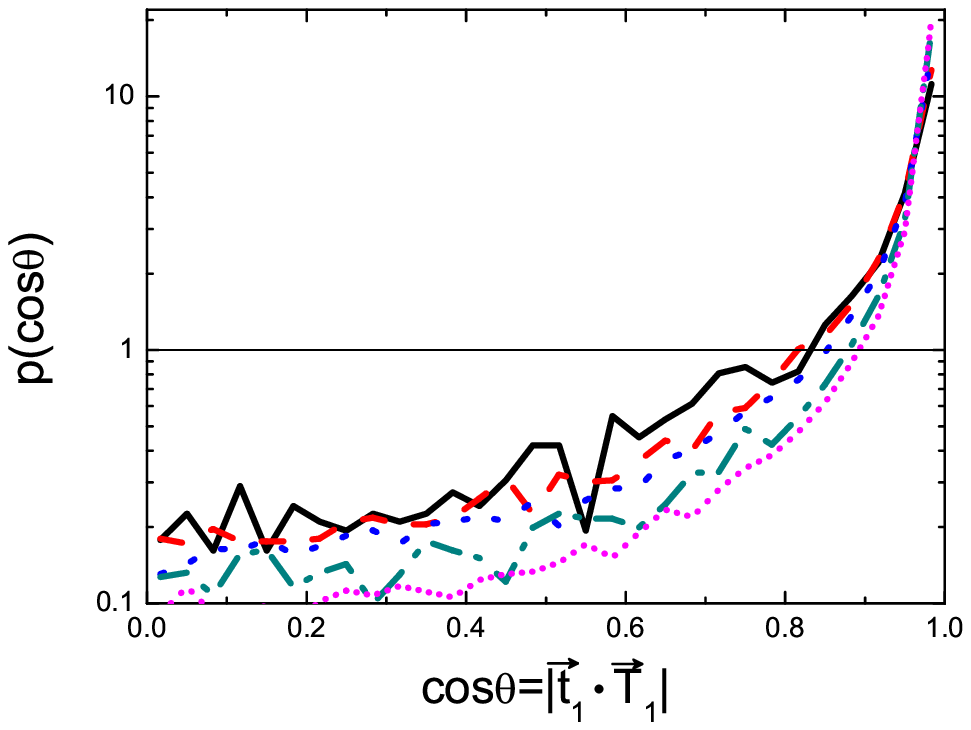} \caption{The distribution
of the cosine of the angle between the mass tidal field
$\vec{T}_1$ and the halo tidal field $\vec{t}_1$. Adaptive SMS of
$M_h$ is used to calculate $\vec{T}_1$. The results are shown for
halos in the same five mass bins as shown in Fig.\,\ref{fig_zfhp}.
The horizontal line indicates a random distribution.}
\label{fig_tta}
\end{figure}

The halo tidal field used as our primary environment quantity is
calculated using only part of the mass in the cosmic density
field. In order to see how it is related to the mass tidal field,
we make comparison between these two quantities. Note that the
mass tidal field is computed based on the density field smoothed
with an adaptive SMS of $M_h$. In Fig. \ref{fig_tta} we show the
distribution of the cosine of the angle, $\theta$, between the
major axes of the mass tidal field, $\vec{T}_1$, and the halo
tidal field $\vec{t}_1$. The distribution is strongly peaked near
$\cos\theta=1$, indicating that the orientation of the two tidal
fields are strongly correlated. We note that the alignments for
the other two axes are similar. Since $T_k$ is the partial
differentiation of the gravitational acceleration along the
direction $\vec{T}_k$ at the center of a halo, the mass tidal
force on the halo surface, in the direction $\vec{T}_k$, is $\sim
R_h T_k$, where $R_h$ is the virial radius of the halo in
question.

\begin{figure}
\epsscale{1}\plotone{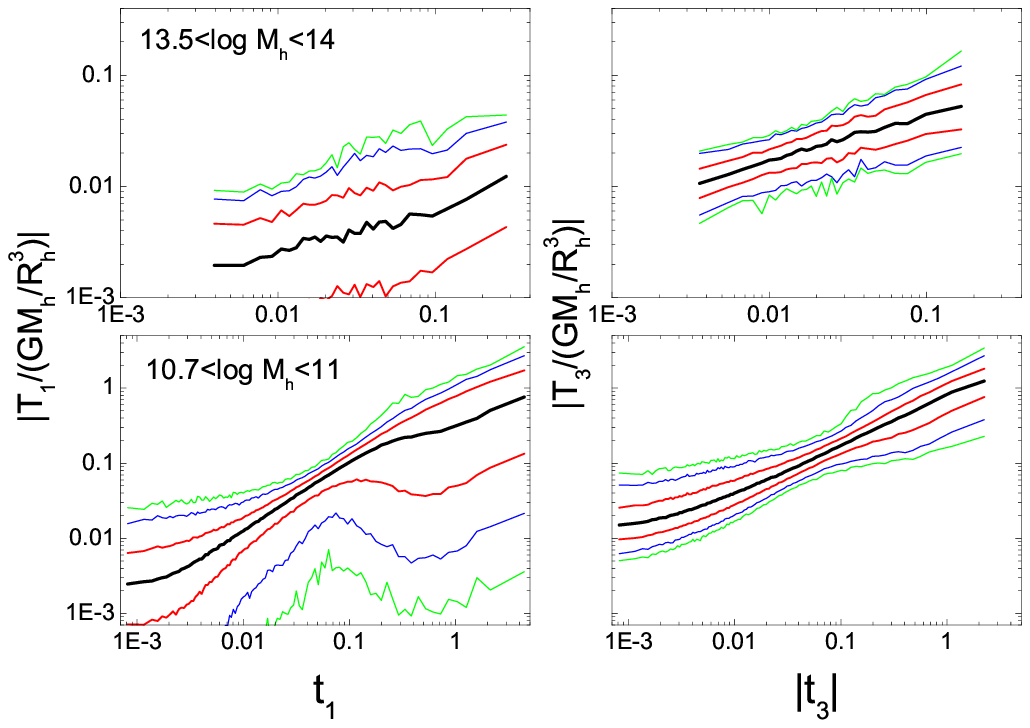} \caption{The comparison between
the mass tidal field $T_k$ and the halo tidal field $t_k (k=1,3)$.
The latter is calculated using halos with $13.5<log(M_h/\msun)
<14$ (upper two panels) and with $10.7<log (M_h/\msun)<11$ (lower
two panels). Adaptive SMS of $M_h$ is used to calculate $T_k$. The
black line represents median, while the colored lines are
percentiles corresponding to $1\sigma$ (68 per cent, red),
$2\sigma$ (95 per cent, blue) and $3\sigma$(99 per cent, green).
Note that the lower blue and green lines in the upper-left panel
are beyond the scope of the plot, and not plotted.} \label{fig_tt}
\end{figure}

In Fig. \ref{fig_tt} we show the absolute value of $R_h
T_k/(GM_h/R_h^2)=T_k/(GM_h/R_h^3)$ versus the absolute value of
$t_k$ for halos in two mass ranges. As one can see, the
eigenvalues of the mass tidal field are strongly correlated with
those of halo tidal field, particularly for low-mass halos. The
only exception is the $|T_1|$ - $|t_1|$ correlation for massive
halos, where the scatter is relatively large.  One possible reason
for this is the contribution of halo self-gravity, which is
included in the mass tidal field, but not in the halo tidal field.
In order to test this possibility, we have made calculations of
the mass tidal tensor with the contribution of the halo's
self-gravity subtracted. We found that halo tidal field match this
external mass tidal field better.

\begin{figure}
\epsscale{1}\plotone{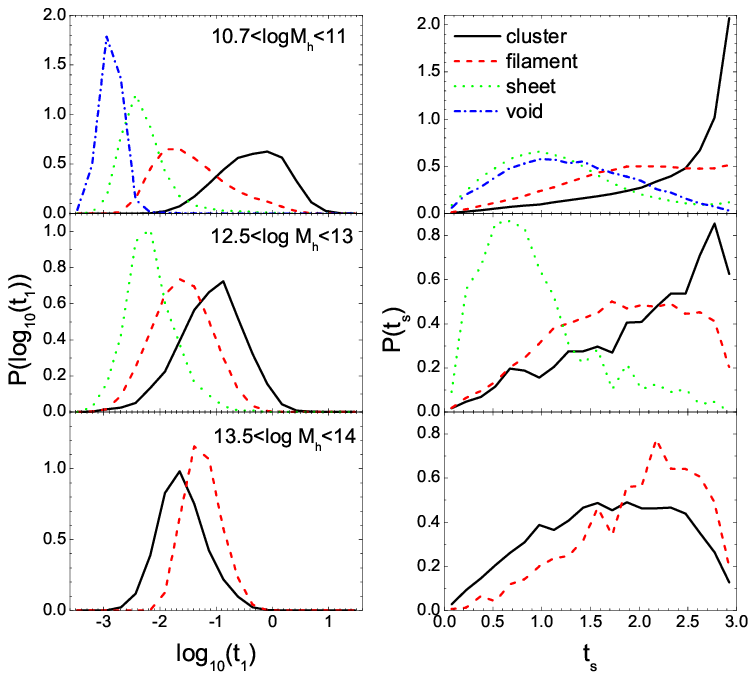}
 \caption{The probability distribution of $t_1$ (left panels)
and $t_s$ (right panels) of halos in three mass ranges and in
various types of environment: cluster, filament, sheet and void
(see the text for the exact definitions). For the most massive
bin, the void population is too small to give a reliable result.}
\label{fig_st1s}
\end{figure}

Overall, the above results demonstrate that the halo tidal field,
which can be estimated from observation, is a good approximation
of the large-scale tidal field produced by the mass density field.
Fig.\ref{fig_st1s} shows the distribution of $t_1$ and $t_s$ for
halos in four different environments, as defined by the signatures
of the eigenvalues of the mass tidal tensor with fixed SMS of
$2M_{\ast}$. Results are shown for halos in three mass ranges,
$10.7<\log (M_h/\msun)<11$, $12.5<\log (M_h/\msun)<13$ and
$13.5<\log(M_h/\msun)<14$. For the two low-mass samples, on
average both $t_1$ and $t_s$ decrease as the type of environment
changes from clusters, to filaments, to sheets and to voids. In
particular, the $t_s$ distribution in cluster regions peaks at 3,
indicating that the tidal field around them is dominated by a
single halo. The $t_s$ distribution in sheet regions peaks at less
than one, suggesting that halos reside in a pancake-like region.
For massive halos of $13.5<\log(M_h/\msun)<14$, however, the
dependence of the distributions on the type of environment is
reversed. Note that the mass tidal field ($T_{k}$) used to
classify the environment is smoothed on a fixed SMS of $2M_\ast$.
If the external tidal field is low for a halo of $M_h > 2M_\ast$,
the tidal field is dominated by the halo's self-gravity, which is
rounder than the external tide. Consequently, $T_1 \sim T_2 \sim
T_3 \sim 4\pi G \rho_m \delta/3>0$, and the halo will be
classified as one residing in cluster. If a halo is located at
high external tidal field region, in which $T_{k}$ is contributed
by both self gravity and large scale environment, it may be
incorrectly identified as a filament halo. Thus, for more massive
halos, where self-gravity contributes more to the mass tidal
tensor, the classification based on the signs of $T_1$, $T_2$ and
$T_3$ may fail to provide a useful description of the real
environment \emph{for this halo}. But such classification may
still be useful \emph{for a test particle} close to the massive
halo.

\begin{figure}
\epsscale{1}\plotone{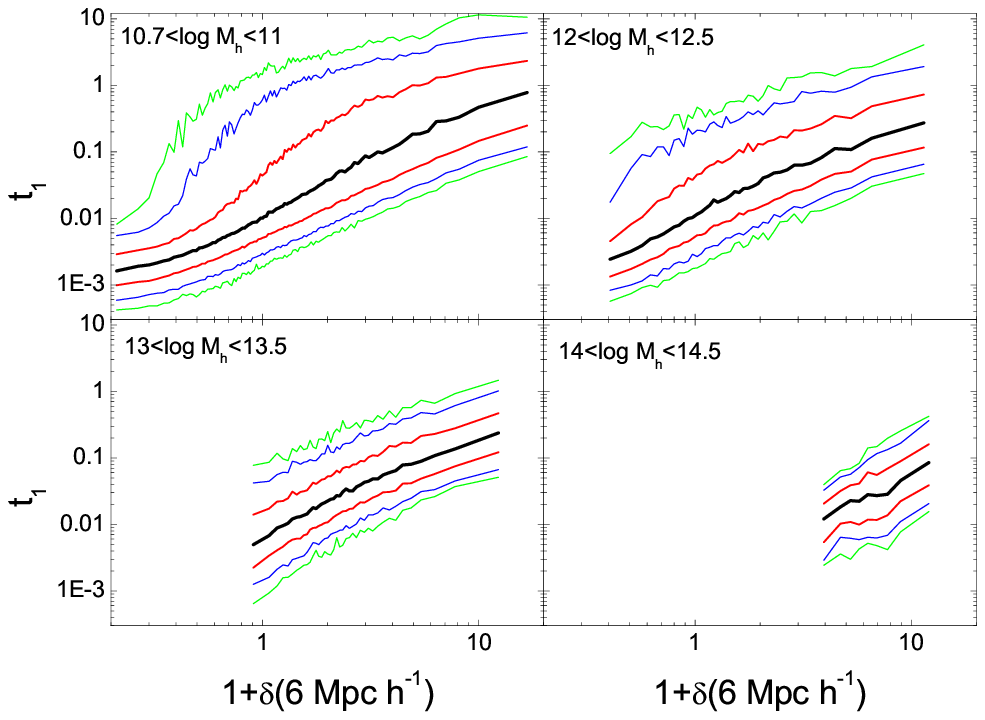}
 \caption{Local overdensity $\delta(6\mpc)$ versus halo tidal
field $t_1$ for halos of various masses. The black line represents
median, while the colored lines are percentiles corresponding to
$1\sigma$ (68 per cent, red), $2\sigma$ (95 per cent, blue) and
$3\sigma$(99 per cent, green).} \label{fig_ovdt1}
\end{figure}

Finally, Fig. \ref{fig_ovdt1} shows $\delta(6\mpc)$ versus $t_1$
for four halo masses. As expected, the halo tidal force on average
increase with the local overdensity. However, there is
considerable scatter in the relation, especially for low-mass
halos. Our tests showed that the large scatter is insensitive to
the choice of $M_{\rm th}$ and the radius for computing the local
overdensity. Some small halos in underdense region, i.e.
$\delta(6\mpc)<0$, can suffer from tidal effects that are
comparable to those in dense region. These small halos may be
close to structures that are much more massive than themselves.

\end{document}